\documentclass[runningheads]{llncs}
\usepackage[T1]{fontenc}
\usepackage{graphicx}
\makeatother
\usepackage[hidelinks, bookmarksopen=true]{hyperref} 
\usepackage{bookmark}

\PassOptionsToPackage{prologue,dvipsnames,table}{xcolor}
\usepackage{amsfonts} 
\usepackage{appendix}
\usepackage{xcolor}
\usepackage{siunitx}
\usepackage{adjustbox,lipsum}
\usepackage{array} 
\usepackage{scalefnt}
\usepackage[skip=0.333\baselineskip]{caption}
\usepackage{pgfplots}
\usepackage{graphicx,booktabs}
\usepackage{footnote}
\makesavenoteenv{tabular}
\makesavenoteenv{table} 
\usepackage{comment}
\usetikzlibrary{calc}
\usepackage{rotating}
\usepackage{makecell}

\usepackage{paralist}
\usepackage{threeparttable}
\usepackage{amsmath}
\usepackage{tabularx}
\usepackage{url}
\usepackage{cleveref}
\usepackage{multirow}
\usepackage{tabularray}
\usepackage{breakcites}
\usepackage{subcaption} 

\let\llncssubparagraph\subparagraph
\let\subparagraph\paragraph
\usepackage{titlesec}
\titlespacing*{\subsubsection}{0pt}{0pt}{0.3em} 
\let\subparagraph\llncssubparagraph

\usetikzlibrary{arrows.meta}
\pgfplotsset{width=13cm,height= 8cm, compat=1.3}
\definecolor{light-gray}{gray}{0.75}
\definecolor{lightest-gray}{gray}{0.85}

\newcolumntype{Y}{>{\centering\arraybackslash}X}

\newcommand{\eg}{e.g., }
\newcommand{\ie}{i.e., }
\newcommand{\wrt}{w.r.t. }
\newcommand{\etal}{\textit{et al. }}

\begin{document}

\title{SoK: 
Measuring Blockchain Decentralization} 


\author{
    Christina Ovezik
    \inst{1}
\and
    Dimitris Karakostas
    \inst{1}
\and
    Mary Milad
    \inst{1}
\and \\
    Aggelos Kiayias
    \inst{1,2}
\and
    Daniel W. Woods
    \inst{1}
}
\authorrunning{C. Ovezik \etal}

\institute{University of Edinburgh \and
IOG \\
}

\maketitle   

\pagestyle{plain}

\begin{abstract}

In the context of blockchain systems, the importance of decentralization is undermined by the lack of a widely accepted methodology to measure it. 
To address this gap, we set out a systematization effort targeting the decentralization measurement workflow. 
To facilitate our systematization, we put forth a framework that categorizes all measurement techniques used in previous work based on the resource they target, the methods they use to extract resource allocation, and the functions they apply to produce the final measurements.
We complement this framework with an empirical analysis designed to evaluate whether the various pre-processing steps and metrics used in prior work capture the same underlying concept of decentralization. 
Our analysis brings about a number of novel insights and observations. 
First, the seemingly innocuous choices performed during data extraction, such as the size of estimation windows or the application of thresholds that affect the resource distribution, have important repercussions when calculating the level of decentralization. 
Second, exploratory factor analysis suggests that in Proof-of-Work (PoW) blockchains, participation on the consensus layer is not correlated with decentralization, but rather captures a distinct signal, unlike in Proof-of-Stake (PoS) systems, where the different metrics align under a single factor.
These findings challenge the long-held assumption within the blockchain community that higher participation drives higher decentralization. 
Finally, we combine the results of our empirical analysis with first-principles reasoning to derive practical recommendations for researchers that set out to measure blockchain decentralization, and we further systematize the existing literature in line with these recommendations.
%

\end{abstract}

\section{Introduction}
The security, reliability, fairness, and even legal status of blockchain systems is
determined in part by their level of decentralization. In practice, decentralization concerns how critical resources, such as tokens or mining power, are distributed across participants \cite{ovezik2024sok}. A system is considered centralized if a small number of parties --- or, in the extreme case, a single party --- controls a large share of the resources that secure the blockchain protocol.
Prospective users or developers of blockchain systems may want to monitor centralization in block production, as it increases the risk of attacks, like double spending or censorship.
Similarly, centralization of token ownership can increase the potential for market manipulation.
Regulators also evaluate decentralization levels to treat blockchain-based assets and classify them as securities or commodities~\cite{securities2019framework,board2019decentralised,mica}.
Thus, decentralization metrics can be used by a wide range of participants across the blockchain
ecosystem, to make quantitative assessments and comparisons of blockchain systems across different layers and over time.

Despite the fundamental importance of decentralization in blockchain systems, there is no widely accepted method for how to measure it in the blockchain and distributed systems literature.
\footnote{An illustratory example can be found in the proceedings of 
the SEC vs. Ripple case, where the SEC argued that the XRP token was centralized and hence a security, while Ripple argued the opposite. Ultimately, the lack of a widely accepted methodology for assessing decentralization led to Ripple being exonerated from most accusations~\cite{sec_v_ripple_2020}. (NB: this ruling is currently under appeal by the SEC).}
This gap in the literature motivated us to undertake a systematization of knowledge on the topic of measuring blockchain decentralization.
To conduct our systematization, we put forward a framework for standardizing decentralization measurements.
This involves addressing three critical questions:
(1) What is the resource of interest, whose allocation one should measure?
(2) How can its distribution be extracted from the available information?
(3) Which functions should be applied to the distribution to produce the final decentralization measurement?

Regarding question (1), the existing literature focuses predominantly on the distribution of protocol participating resources (e.g., hashing power or stake) or the distribution of tokens to different participants.

In terms of question (2), there is a wide variety of approaches that demand careful systematization.
One key issue is that the distribution of resources may not lend itself immediately to extraction.
For instance, the amount of resources, such as computational power, held by consensus participants cannot be observed directly.
Instead, block production is typically used as a proxy for it, but this approach requires caution --- for instance,
inferring hashing power from the number of produced blocks involves fixing an estimation window, the choice of which can significantly influence the outcome of the analysis.
Other considerations include how to attribute blocks or tokens to entities and whether to prune the data set of token holders to approximate the most useful segment of the token ownership distribution, given the pseudonymity of addresses.

Finally, regarding question (3), a plethora of functions have been put to the test in the literature. Given the overlapping nature of many of these functions, the interesting part from a systematization perspective is to identify what exactly they measure and to what extent their estimation truly captures decentralization.
\looseness=-1 

\subsubsection*{Contributions.}
Our first contribution is the development of a framework that describes the core workflow of blockchain decentralization measurements and facilitates our systematization. This includes: (1) choice of resource; (2) data pre-processing, and in particular choices on (2a) clustering, (2b) resource estimation, (2c) population estimation, (2d) measurement frequency, and (2e) inclusion thresholds; and (3) choice of decentralization metric.

We then systematize the blockchain decentralization literature using the above framework, to showcase how prior work has approached the various steps of the measurement workflow.
We observe that the literature lacks a unified methodology for addressing these concerns, with studies making a wide variety of choices, frequently without complete justification. 

We complement our systematization with an empirical analysis using data from live blockchain systems, demonstrating that different methodological choices taken in the literature have a significant impact on decentralization estimates.
In particular, our analysis reveals that the results can be manipulated via arbitrary decisions on how to pre-process data, such as removing wallets with balance below a certain threshold, or aggregating block data over an arbitrary time period.

We  review all decentralization metrics from prior work and perform exploratory factor analysis to uncover the underlying constructs they capture.
A key finding is that participation, namely the number of entities that actively command resources, and decentralization, namely how such entities share these resources, do not always align, with distinct factors emerging to represent the two in some cases.
This challenges the common assumption that high participation inherently leads to high decentralization in blockchain systems.

Based on our systematization and empirical analysis, we derive a range of recommendations for researchers who set out to measure decentralization in blockchain systems and organize the existing literature along these recommendations.

\subsubsection*{Scope.} Our systematization framework applies to fungible resources distributed across entities that participate in a blockchain system.
The canonical applications are tokens distributed across users, and computational power or stake distributed across block producers.
The framework can also extend to layer-2 blockchains or DeFi applications, such as those involving on-chain governance tokens. 
However, it is not suitable for blockchain layers involving non-fungible resources, such as the network layer, where the graph structure must be incorporated into any analysis of decentralization.
We further note that our pragmatic approach differs from prior systematizations that have focused on conceptualizing decentralization~\cite{ovezik2024sok,zhang2023sok}.

\subsubsection*{Roadmap.}
The remainder of this paper is organized as follows. Section~\ref{sec:framework} describes a framework that identifies the decisions that need to be made to measure decentralization.
Section~\ref{sec:relatedwork} describes our methodology to systematize the existing literature 
and classifies prior work based on our framework.
Section~\ref{sec:empirical} presents our own empirical analysis that highlights how arbitrary decisions can manipulate estimates of decentralization.
Finally, Section~\ref{sec:recommendations}
derives recommendations for future measurements, with the overarching goal of providing standardized, architecture agnostic estimates of decentralization at the consensus and tokenomics layers and describes the existing literature vis-\`a-vis those recommendations. 
\section{Systematization framework}
\label{sec:framework}

Our first contribution is a systematization of the decisions that researchers need to make when measuring decentralization in blockchain systems.
These decisions are:
(1) determining which data to collect (layer / resource);
(2) pre-processing collected data, which involves choices about
(2a) clustering,
(2b) resource estimation windows,
(2c) population estimation,
(2d) measurement frequency,
and
(2e) inclusion thresholds; and
(3) choosing which metrics to use to derive insights about decentralization.

Some of these questions are straightforward and have been addressed in prior work, while others are more nuanced and are frequently overlooked during the measurement process. 
In the following paragraphs, we will discuss each step in detail and explain why all are critical for a robust decentralization analysis.

\subsection{Blockchain layers and resources} 

Ideally, a decentralization measurement process would take a blockchain system as input and output a value that describes its overall degree of decentralization.
However, blockchains are multilayered systems and some layers may exhibit higher decentralization than others.
%
This means that any decentralization analysis begins, explicitly or implicitly, by choosing which layer to focus on.\footnote{Developing a unified index to represent the decentralization of all layers has been posed as an open question in the literature~\cite{ovezik2024sok}, yet there is no known methodology to date that achieves this.}

A number of recent research works have been dedicated to identifying the different layers of blockchain systems that are relevant to decentralization~\cite{ovezik2024sok,zhang2023sok,sai2020taxonomy}.
The four layers that are common across all of them are: consensus, tokenomics,\footnote{There is no consistent name for this layer in the literature. We adopt the term ``tokenomics'' from \cite{ovezik2024sok}, but it has also been referred to as ``wealth''~\cite{zhang2023sok} or ``incentives''~\cite{sai2020taxonomy}.} network, and governance.

The consensus layer captures direct participation in the protocol, which happens through the creation of blocks that extend the ledger.
This layer carries direct security implications --- 
for example, if an entity controls the majority of block production, it can launch a ``51\% attack'' and double-spend coins~\cite{aponte202151}. 
 
The tokenomics layer focuses on the distribution of wealth within a system. 
This layer is strongly tied to the utility of the system and is also particularly important from a policymaking perspective, with critical decisions depending on the structure of token ownership. For instance, a recent U.S. bill proposes classifying a blockchain asset as a security if there exists a single entity that controls more than 20\% of its total token supply~\cite{congress-bill}.

The network layer concerns the peer-to-peer protocol underpinning the blockchain, focusing on aspects such as node distribution and connectivity.
Decentralization on this layer is critical for ensuring that information can propagate effectively and without being censored.
 
The governance layer encompasses all decision-making activity that is relevant to a blockchain, which can take place off-chain, \eg through forums or traditional corporate structures, or on-chain, using mechanisms that assign voting rights to tokens. 

Out of these layers, consensus and tokenomics have received by far the most attention in the relevant literature, with numerous studies attempting to measure the decentralization of block production or wealth in popular blockchains.
In contrast, there are only a handful of studies that address the decentralization of the network layer~\cite{gencer2018decentralization,gochhayat2020measuring,howell2023measuring}.
Governance of layer-1 blockchains has also received very little attention in decentralization studies~\cite{azouvi2019egalitarian,Jia}, though recently some studies have explored the  governance decentralization of DeFi projects~\cite{barbereau2022defi,jensen2021decentralized,sun2023decentralization}.
Notably, in systems that employ on-chain governance, voting rights are represented by tokens, hence the methodology described for the tokenomics layer can be applied to measure decentralization on the governance layer, too, including that of DeFi applications. 
Nonetheless, given the limited research on these layers, we consider it premature to systematize the decentralization measurement process for network and governance.
Instead, we focus our systematization efforts on the consensus and tokenomics layers from now on.

A further benefit of focusing on consensus and tokenomics is that these layers are consistent across different systems, as all blockchains generate blocks and issue tokens.
%
%
Moreover, consensus and tokenomics data can be obtained directly or inferred from the blockchain itself, allowing for transparent data processing and analysis.
\looseness=-1

To assess the decentralization of a given layer, one must first identify the layer's key resource and determine how it is distributed among participating entities~\cite{ovezik2024sok}.
This process is straightforward in some cases --- for example, the tokenomics layer can be assessed by examining the distribution of the system's native token.
In some layers, however, the ``ideal'' resource is not directly observable, thus necessitating the use of a ``proxy'' resource. This is true for the consensus layer, where the goal is to measure how the \emph{power} to create blocks is distributed, yet only the outcomes of the block creation process can be observed.

\subsection{Data pre-processing}
The process of turning arbitrary, possibly complex and semantically related,
data into an element of the domain of a decentralization metric function introduces various
interesting pre-processing choices. In particular, these steps include
clustering addresses controlled by the same entity,  estimating distributions that are not directly observable, potentially pruning these distributions, and capturing snapshots of them at different points in time.
In the following, we explore the steps required to transform
raw ledger data into distributions that can form the input of the metric functions.

\subsubsection{Clustering of blocks and addresses.}
Clustering is the process of linking all resources associated with a single real-world entity. It is a crucial step in a decentralization analysis, as unclustered data can yield misleading decentralization estimates.
At the consensus layer, the goal is to attribute all blocks to the real-world entity that created them, such as a mining pool.
At the tokenomics layer, clustering involves grouping seemingly unlinked addresses under the entity that controls them.
\looseness=-1 

Notably, this step hinges on defining what constitutes an ``entity'', which can be a contestable subject in itself. 
For instance, in the consensus layer, a mining pool could be treated as a single entity or as a collection of individual miners, each regarded as a separate entity. 
In the tokenomics layer, similar considerations arise when dealing with custodial wallets, shared accounts, or corporations that have multiple subsidiaries. 
In line with prior work on characterizing entities in blockchain ecosystems, we consider all organizations, such as exchanges or mining pools, to be distinct entities
\cite{DBLP:conf/icdm/JourdanBWD18}.

In UTxO-based ledgers, such as Bitcoin, clustering is particularly important, as the system's design inherently drives users to have multiple addresses. 
Over the years, various heuristics have been proposed to cluster these addresses based on the transaction graph.
One such heuristic takes advantage of transactions that consume the assets of multiple addresses~\cite{DBLP:conf/socialcom/ReidH11}. 
Intuitively, the ability to spend assets from multiple
addresses suggests a direct link between these addresses, \eg that one wallet controls all of them. 
This heuristic is the oldest and most widely
applied~\cite{DBLP:conf/socialcom/ReidH11,DBLP:conf/icdm/JourdanBWD18,DBLP:conf/fc/RonS13,DBLP:conf/uic/HarriganF16,DBLP:journals/cacm/MeiklejohnPJLMV16}.
Another heuristic involves identifying a transaction's ``change'' address and linking it to the input addresses~\cite{DBLP:conf/fc/AndroulakiKRSC13}. 
For the case of Cardano, an additional heuristic has been developed, which leverages its on-chain delegation mechanism and groups together addresses that share the same ``stake key'', since this key can be used to withdraw rewards~\cite{chegenizadeh2023heuristic}.

In account-based blockchains, like Ethereum, addresses are typically reused, making clustering less critical compared to UTxO-based systems. 
Nonetheless, clustering can still be useful for identifying entities that control multiple addresses.
Among the proposed heuristics for these blockchains, the most effective involves clustering addresses that receive funds from the same exchange-generated deposit address~\cite{victor2020address}.

Finally, a crowd-sourced method for enhancing the clustered information is ``tagging''. 
In this case, researchers use public information from blockchain explorers,
social media, or other sources to attribute addresses to their owners. 
The effectiveness of tagging is difficult to assess, since the information is
often manually gathered or provided by proprietary means.
Alternatively, self-tagging allows parties to voluntarily disclose ownership of resources.
Although this is rare on the tokenomics layer, it is prevalent on the consensus layer, where mining pools often announce a block's production by including their name within it, effectively advertising their operation.

\subsubsection{Resource estimation window.}
In some cases, the distribution of resources can be directly
observed from the ledger. For example, the token balance of each address is publicly available and updated in real time. In other cases, like the consensus
layer of Proof-of-Work (PoW) blockchains, this is not as straightforward.
The resource of interest there is the
computational (or mining) power of each block producer. Since this power cannot be observed directly, it must be estimated, typically using produced blocks as a
proxy. 
Specifically, when looking at a window of $n$ blocks, an entity that creates $x$ of the observed blocks is assumed to control a fraction $\frac{x}{n}$ of the total mining power.
%
%
A critical consideration here is the trade-off between statistical confidence
and time-specificity. 
A larger sample size increases statistical confidence but may overlook temporal changes. 
On the other hand, using only few blocks, \eg produced within a single day, provides high time-specificity but lacks statistical power to draw reliable conclusions~\cite{milad2024statistical}.

For Proof-of-Stake (PoS) blockhains, similar considerations arise.
Even though the stake distribution can be observed from the blockchain, not all PoS protocols use this exact distribution in their consensus process. 
For example, some systems employ a periodic sub-selection of parties to run the consensus protocol, which only  reflects a small part of the stake distribution.\footnote{An example of this is Cosmos, where only the top X validators (currently X = 180) are selected to participate in consensus, while all the rest are considered inactive~\cite{kwon2019cosmos}.}
Therefore, since the resource of interest in consensus is protocol-participating power, extracting this from the stake distribution would need to take into
account protocol-specific details, such as how each protocol implements the sub-selection mechanism.
To avoid complexities specific to each system and remain agnostic to PoS architectures, researchers may instead apply the same approach as for PoW systems, using block production to estimate the consensus power distribution.

\subsubsection{Population estimation.}
A related but distinct challenge is estimating the total population of active participants at a given time.
Since mining is a probabilistic process, miners may fail to produce
\emph{any} blocks during specific time periods, even if they contribute computational power.
If the population of active participants is not estimated separately, unsuccessful miners may be excluded, skewing metrics that rely on the full distribution.

A simple method is to use a broader observation window than the resource estimation window, considering an entity active if it produces at least one block within this larger timeframe. 
In general, increasing the size of the population estimation window will better capture unsuccessful miners, but risks capturing entities who are no longer active.
Another approach is to track the oldest and most recent blocks produced by each miner and assume they were active during the intervening period. 
A more sophisticated approach could involve analyzing block production patterns to perform statistical inference.

Ideally, the population estimation method should include all entities contributing computational power during the resource estimation window,
but exclude those that participated before or after but were inactive
during this period.

\subsubsection{Measurement frequency.}
Another choice to make during a decentralization analysis is how often to sample the data.
Analyzing a snapshot of the system requires only a single data point, so it is easier to measure. 
However, conducting a longitudinal analysis can capture
trends over time or assess how specific events impact the system's decentralization levels.
In such historical analyses, collecting enough measurement points is crucial to provide a granular view of how the system evolved over time. 
If the measurement frequency is too low, important shifts or fluctuations in decentralization could be missed.
%

Notably, the measurement frequency is distinct from the resource and population estimation windows. 
While estimation windows impact how resources or
participants are distributed within a timeframe, measurement frequency determines how often those windows are sampled. 

\subsubsection{Inclusion thresholds.}
Finally, researchers must decide whether to analyze the entire distribution or restrict it by 
applying inclusion criteria.
This is particularly relevant for the tokenomics layer, where token distribution typically exhibits a long tail, with thousands or millions of addresses controlling tiny amounts of assets, \eg ``dust'' wallets~\cite{perez2019another}.
Therefore, it may be reasonable to consider some of these addresses inactive and exclude them from the analysis, to avoid skewing metrics that are sensitive to the tail of the distribution. 
%
%
Importantly, while other pre-processing steps can be performed in parallel, inclusion criteria should always be applied after clustering and resource / population estimation.


\subsection{Decentralization metrics}
A decentralization metric is a function $M: \mathbb{R}^* \to \mathbb{R}$, where the domain is a distribution of fungible resources across a number of parties and the output is a real number.
Various metrics have been used over the years to measure the decentralization of blockchain systems, either crafted for them, or borrowed from fields like economics and information theory.
Here, we focus on the metrics that have been used at least twice in the blockchain decentralization literature.

\noindent\textbf{Shannon Entropy.} 
Shannon entropy, also known as information entropy, measures the average degree of ``unpredictability'' inherent to the outcomes of a random variable~\cite{shannon1948mathematical}.
Formally, it is defined as:\\
${\mathrm{H} (X):=-\sum_{x\in {\mathcal{X}}}p(x)\log p(x)}$, 
where $X$ is a discrete random variable that takes values in $\mathcal{X}$ and $p(x)$ is the probability of an outcome $x 
\in \mathcal{X}$. 
Higher entropy corresponds to greater uncertainty (\ie smaller probabilities for each possible value), while lower entropy reflects more predictable outcomes. 

In the context of blockchains, if the random variable represents block creators and the block distribution is treated as a probability distribution over possible outcomes, entropy reveals how difficult it is to predict who creates blocks. 
Intuitively, higher entropy values indicate greater decentralization~\cite{wu2019information}.
To take an extreme example, in a highly centralized setting where all blocks are produced by a single entity, entropy would be 0, as there is no uncertainty regarding who may create some block.

\noindent\textbf{Gini Coefficient.}
The Gini coefficient is a widely used measure of inequality, typically applied in economics to quantify income or wealth disparities~\cite{gini1921measurement,gini-world-bank}. 
%
To define Gini, we consider two curves. First, the Lorenz curve, which plots
the proportion of the total assets ($y$-axis) that are cumulatively controlled
by the bottom $x$ of entities. Second,  the line of equality, \ie the line at
$45$ degrees starting at $(0, 0)$. The Gini coefficient is the ratio of the
area between the Lorenz curve and the line of equality over the total area
under the line of equality:
$G = \frac{A-B}{A}, \ \text{where} \ A = \frac{n\sum_{i=1}^nx_i}{2} \ \text{and} \ B = \sum_{i=1}^nix_i$.

A Gini value of $0$ represents perfect
equality, whereas $1$ indicates complete centralization of assets to a single
entity. 
Notably, the Gini coefficient takes into account the entire distribution, including zero values.
For example, the Gini of $X = \{3, 2, 1\}$ differs significantly from that of  $Y = \{3, 2, 1, 0\}$ ($0.22$ vs $0.42$).
Therefore, to obtain accurate Gini values, it is essential to capture a complete view of resource distribution among parties, \eg including block producers who attempted to produce blocks within a given time period but failed.
A closely related metric to Gini is the Theil index, which intends to capture the lack of diversity, or the redundancy, in a population~\cite{conceiccao2000young}.
While the Theil index has been discussed in the blockchain decentralization literature~\cite{mueller2022understanding}, it has never been measured empirically.

\noindent\textbf{Nakamoto Coefficient.}
The Nakamoto coefficient
(NC) represents the minimum number of parties that collectively 
control a majority of resources~\cite{nakamotoCoefficient}. This metric was introduced in the 
context of blockchains 
and is commonly used in blockchain analysis, particularly when examining block production data. 
In a distribution with $n$ entities in which  entity $i$ controls fraction $s_i$ of the total resources, the Nakamoto coefficient is defined as: 
$\text{NC} = min\{k \in [1,...,n] : \sum_{i=1}^k s_i > 0.5\}$.
Notably, some research works, \eg \cite{lin2021measuring}, use $\geq 0.51$ instead of $> 0.5$ in their definition of the Nakamoto coefficient. 
We attribute this to the misnomer of the ``$51\%$ attack'', which may lead people to believe that $51\%$ of resources are needed to compromise the security of a ledger, when in fact a majority ($> 50\%$) suffices~\cite{li2020survey}.

The NC can be particularly useful when analyzing
decentralization, as it identifies the minimum number of parties that need to
collude, in order to gain a majority of the resources under question. This 
quantity is highly relevant when arguing about a distributed ledger's security, 
which often relies on the assumption that a majority of power is honest, \eg in the case 
of Bitcoin (PoW)~\cite{EC:GarKiaLeo15} or Ouroboros 
(PoS)~\cite{C:KRDO17}.

\noindent\textbf{$\tau$-decentralization Index.}
An intuitive generalization of the Nakamoto coefficient is the number of entities that collectively control more than any $\tau$ fraction of the total resources. Formally, this is defined as 
$min\{k \in [1,...,n] : \sum_{i=1}^k s_i > \tau\}$, 
where $n$ is the total number of entities in the system and $s_i$ is the share of resources controlled by entity $i$.
Different values of $\tau$ may be interesting to explore, depending on the context. 
For instance, the liveness threshold of systems like
Algorand~\cite{gilad2017algorand} or Ethereum
PoS~\cite{buterin2020combining,buterin2020incentives} is $33\%$, instead of
$50\%$ so exploring this value of $\tau$ can be particularly useful there. 

Notably, this metric has been used in the literature before, but without a consistent
name. For instance, Nadler \etal use the term ``top $n$\%'' for it and calculate it for the
top $99\%$~\cite{nadler2020decentralized}.
Additionally, $\tau$ is similar to the $N_\epsilon$ ``centralization level'', which states that $N$ parties control $1-\epsilon$ of all resources, introduced by Chu \etal~\cite{chu2018curses}. 

\noindent\textbf{Concentration ratios.}
Concentration ratios have long been employed in economics to assess the degree
of market concentration within an industry. For a given number $n$, the
concentration ratio $CR_n$ represents the combined market share of the $n$
largest firms in an industry. The most commonly used concentration ratios in
economic analysis are $CR_3, CR_4$, and $CR_5$~\cite{Shepherd2018}. Formally,
the $n$-concentration ratio of a market is defined as:
$CR_n = \sum_{i=1}^n s_i$, 
where $s_i$ is the market share of the $i^{th}$ largest entity.

In blockchain studies, concentration ratios have been applied several times to
analyze decentralization, though they are commonly called ``mining power ratios''
\cite{gencer2018decentralization}.
We adopt the more neutral term ``concentration ratios'', as it aligns with the
historical usage of the metric and allows for broader application to any
resource, such as the distribution of wealth in the system, rather than solely to mining power.

\noindent\textbf{Herfindahl–Hirschman index (HHI).}
The Herfindahl–Hirschman
index is a market concentration metric, measuring the
size of entities \wrt the industry in which they operate~\cite{hhi}. The HHI is computed 
as the sum of the squares of each entity's percentage of market control: 
$\text{HHI}=\sum _{i=1}^{n}(s_{i})^{2}$, 
where $s_{i}$ is the market share of entity $i$ and $n$ is the number of all
entities. The HHI ranges from $0$ to $10{,}000$, with $0$ indicating
the existence of many small entities and $10{,}000$ indicating a monopoly.
%

Interestingly, because the HHI has been used by antitrust authorities, there
exist guidelines to interpret its results. For instance, US agencies classify
markets in three types, based on the HHI:
i) unconcentrated markets: HHI below $1{,}500$;
ii) moderately concentrated markets: HHI between $1{,}500$ and $2{,}500$;
iii) highly concentrated markets: HHI above $2{,}500$~\cite{united2010horizontal}.
Similar guidelines could be developed for evaluating blockchain decentralization.

\noindent\textbf{Number of parties.}
Finally, a basic method that has been used to estimate decentralization is to count the number of entities
controlling any amount of the resource of interest. 
This approach provides a straightforward quantitative assessment, as a higher number of active participants generally indicates a more decentralized system. However, it is essential to recognize that this metric has its limitations, as it does not account for disparities in power among participants.

\section{Literature Systematization}
\label{sec:relatedwork}

\begin{figure}[t]
    \centering
    \includegraphics[width=\linewidth]{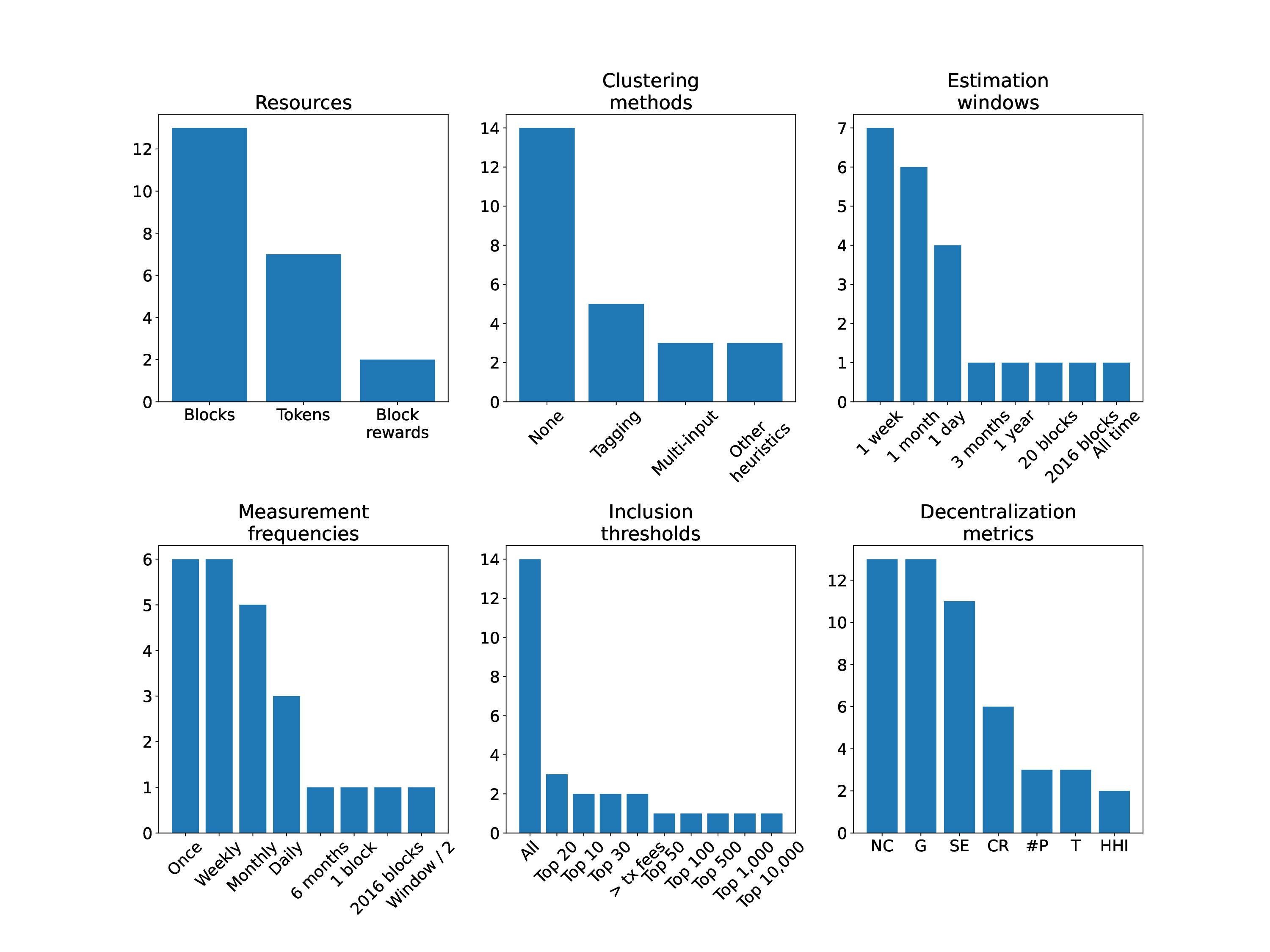}
    \caption{Frequency of each measurement choice in blockchain decentralization literature. Note that resource estimation windows and population estimation windows have been merged to one chart because the literature does not differentiate between the two.}
    \label{fig:histograms}
\end{figure}


Using the framework of Section~\ref{sec:framework}, we now review prior work and evaluate how it addresses the questions outlined above.
Table~\ref{tab:arbitrary-choices} summarizes this classification, organized in chronological order of publication. 

\begin{table}
\scriptsize
\begin{tabular}{|c|c|c|ccc|c|c|}
\hline
\textbf{Ref} & \textbf{Resource} & \textbf{Clustering} & \multicolumn{1}{c|}{\textbf{\begin{tabular}[c]{@{}c@{}}Resource \\ estimation \\ window\end{tabular}}} & \multicolumn{1}{c|}{\textbf{\begin{tabular}[c]{@{}c@{}}Population\\ estimation\end{tabular}}} & \textbf{\begin{tabular}[c]{@{}c@{}}Measurement\\ frequency\end{tabular}} & \textbf{\begin{tabular}[c]{@{}c@{}}Inclusion \\ threshold\end{tabular}} & \textbf{Metrics} \\ \hline
\cite{gencer2018decentralization} & Blocks & Tagging & \multicolumn{3}{c|}{1 Week} & Top 20 & $\mathsf{CR_{1,3,4}}$ \\ \hline
\cite{valdivia2019decentralization} & Blocks & - & \multicolumn{3}{c|}{1 Month} & - & $\mathsf{CR_{3}}$ \\ \hline
\multirow{2}{*}{\cite{wu2019information}} & Blocks & - & \multicolumn{2}{c|}{1 Week} & Once & - & SE \\ \cline{2-8} 
 & Tokens & - & \multicolumn{2}{c|}{N/A} & Once & Top 10,000 & SE \\ \hline
\cite{romiti2019deep} & Blocks & \begin{tabular}[c]{@{}c@{}}Tagging,\\ Multi-input\end{tabular} & \multicolumn{3}{c|}{2,016 Blocks} & - & G, NC \\ \hline
\cite{gochhayat2020measuring} & Blocks & - & \multicolumn{2}{c|}{\begin{tabular}[c]{@{}c@{}}1,3 Months\\ 1 Year \\ All-time\end{tabular}} & Once & - & \begin{tabular}[c]{@{}c@{}}SE, G, \\ NC, \#P\end{tabular} \\ \hline
\cite{wang2020measurement} & Blocks & Tagging & \multicolumn{3}{c|}{1 Day} & - & NC,$\mathsf{CR_{3,5}}$ \\ \hline
\cite{li2021comparison} & Blocks & - & \multicolumn{2}{c|}{1 Month} & Once & Top 10, 20, 30 & \begin{tabular}[c]{@{}c@{}}SE, NC, \\ $\mathsf{CR_{1,2,4}}$, $\mathsf{T_{0.33}}$\end{tabular} \\ \hline
\cite{nadler2020decentralized} & Tokens & \begin{tabular}[c]{@{}c@{}}Other \\ heuristics\end{tabular} & \multicolumn{2}{c|}{N/A} & 1 Month & Top 500 & \begin{tabular}[c]{@{}c@{}}G, NC, \\ $\mathsf{CR_{5,10,50,100,500}}$,\\ $\mathsf{T_{0.99}}$, \#P\end{tabular} \\ \hline
\cite{jensen2021decentralized} & Tokens & - & \multicolumn{2}{c|}{N/A} & Once & > tx fees & G, NC \\ \hline
\cite{lin2021measuring} & Blocks & - & \multicolumn{2}{c|}{\begin{tabular}[c]{@{}c@{}}1 Day\\ 1 Week\\ 1 Month\end{tabular}} & \begin{tabular}[c]{@{}c@{}}1, 0.5 Day\\ 1, 0.5 Week\\ 1, 0.5 Month\end{tabular} & - & SE, G, NC \\ \hline
\cite{zeng2021characterizing} & Rewards & - & \multicolumn{3}{c|}{1 Week} & - & NC, $\mathsf{CR_{3,5}}$ \\ \hline
\cite{kusmierz2022centralized} & Tokens & - & \multicolumn{2}{c|}{N/A} & Unclear & Top 30, 50, 100 & SE, G, NC \\ \hline
\cite{Jia} & Tokens & - & \multicolumn{2}{c|}{N/A} & Once & Top 1,000 & SE, G \\ \hline
\cite{barbereau2022defi}& Tokens & Tagging & \multicolumn{2}{c|}{N/A} & 6 Months & > tx fees & SE, G \\ \hline
\cite{liu2022understanding} & Blocks & - & \multicolumn{3}{c|}{1 Month} & Top 10, 20, - & SE, \#P \\ \hline
\cite{mueller2022understanding} & Blocks & - & \multicolumn{2}{c|}{20 Blocks} & 1 Block & - & G \\ \hline
\multirow{2}{*}{\cite{campajola2023evolution}} & Blocks & Multi-input & \multicolumn{3}{c|}{1 Week} & - & NC \\ \cline{2-8} 
 & Tokens & \begin{tabular}[c]{@{}c@{}}Multi-input,\\ Other \\ heuristics\end{tabular} & \multicolumn{2}{c|}{N/A} & 1 Week & - & G \\ \hline
\cite{grandjean2023ethereum} & Blocks & \begin{tabular}[c]{@{}c@{}}Tagging, \\ Other \\ heuristics\end{tabular} & \multicolumn{2}{c|}{1 Week} & 1 Day & - & \begin{tabular}[c]{@{}c@{}}G, NC, \\ HHI, $\mathsf{T_{0.33}}$\end{tabular} \\ \hline
\cite{liu2023demonitor} & Blocks & - & \multicolumn{3}{c|}{\begin{tabular}[c]{@{}c@{}}1 Day\\ 1 Week\\ 1 Month\end{tabular}} & - & SE, G, NC \\ \hline
\cite{yan2024analyzing} & Rewards & - & \multicolumn{3}{c|}{1 Day} & - & \begin{tabular}[c]{@{}c@{}}SE, G, \\ NC, HHI\end{tabular} \\ \hline
\end{tabular}
\caption{Measurement approaches in blockchain decentralization literature, ordered chronologically. The metrics are: Shannon Entropy (SE), Gini coefficient (G), Nakamoto coefficient (NC), Herfindahl-Hirschman index (HHI), concentration ratios (CR), $\tau$-decentralization index (T), and number of parties (\#P).}
\label{tab:arbitrary-choices}
\end{table}

To compile the articles included in Table~\ref{tab:arbitrary-choices}, we performed a systematic search using three databases: Google Scholar, Semantic Scholar, and DBLP. 
We used the keywords ``measuring blockchain decentralization'' and selected among the results the studies that conducted empirical decentralization measurements. 
As discussed in the previous section, our systematization focuses on layers with fungible resources, so we exclude articles on the network layer.
%
Using this initial sample, we applied a backward and forward reference search (snowballing) to identify additional relevant studies and applied the same selection criteria to them.
\looseness=-1 

Importantly, Table~\ref{tab:arbitrary-choices} includes only measurement studies, that is, works that analyze data from real-world blockchain systems to derive insights about their decentralization.
Our work is also informed by complementary literature, such as systematizations of blockchain decentralization dimensions~\cite{ovezik2024sok,zhang2023sok,sai2020taxonomy}, game-theoretic treatments of decentralization~\cite{bahrani2024centralization}, simulation-based analyses~\cite{ovezik2022decentralization}, or studies that propose (de)centralization indices without conducting direct measurements\cite{chu2018curses}.
However, these contributions are orthogonal to 
our systematization and were therefore excluded from the table.

In cases where certain decisions were not explicitly discussed in some paper, we made
reasonable assumptions to complete the table. 
Specifically, since none of the studies discussed population estimation as a distinct decision, we assumed it aligned with the resource estimation window. 
Also, if clustering was not explicitly mentioned, we assumed that
no clustering was employed. 

Table~\ref{tab:arbitrary-choices} reveals that there is no standardized approach in the literature for performing the steps identified in our framework, undermining one's ability to compare results across studies.
This is also highlighted in Figure \ref{fig:histograms}, which depicts the frequency of different options employed at the various stages of the decentralization measurement workflow.
In the following paragraphs, we elaborate on how each step of the framework has been addressed in prior research.

\subsubsection{Layers and resources.} \label{subsec:layers}
A number of studies measure decentralization on the
consensus~\cite{gencer2018decentralization,li2021comparison,lin2021measuring,wu2019information,liu2022understanding,liu2023demonitor,mueller2022understanding,campajola2023evolution,grandjean2023ethereum}
and tokenomics~\cite{DBLP:conf/fc/RonS13,campajola2023evolution,nadler2020decentralized,wu2019information,sai2021characterizing,kondor2014rich,kusmierz2022centralized,Jia,cheng2021decentralization}
layers. 
For the consensus layer, most works measure the distribution of produced blocks, with the exception of two studies~\cite{zeng2021characterizing,yan2024analyzing}
that instead analyze the distribution of rewards among block producers. 
For the tokenomics layer, all studies used the system's native token as the resource of interest.


\subsubsection{Clustering.}
Notably, the bulk of reviewed studies did not use any clustering techniques to
link resources controlled by a single entity~\cite{valdivia2019decentralization,wu2019information,gochhayat2020measuring,li2021comparison,jensen2021decentralized,lin2021measuring,zeng2021characterizing,kusmierz2022centralized,Jia,liu2022understanding,mueller2022understanding,liu2023demonitor,yan2024analyzing}.
This omission likely results in an overestimation of decentralization in their findings. 
Among the studies that did apply clustering, some relied on blockchain explorers for resource attribution through tagging~\cite{gencer2018decentralization,romiti2019deep,wang2020measurement,barbereau2022defi,grandjean2023ethereum}, 
whereas others used primary heuristics~\cite{romiti2019deep,nadler2020decentralized,campajola2023evolution,grandjean2023ethereum}, such as multi-input transactions. 

\subsubsection{Resource estimation window.}
The two broad approaches for the resource estimation window are time-based
windows and block-based windows. Most commonly a time-based window is employed,
with most studies opting for windows like one
day~\cite{wang2020measurement,lin2021measuring,liu2023demonitor,yan2024analyzing}, one
week~\cite{gencer2018decentralization,wu2019information,lin2021measuring,zeng2021characterizing,campajola2023evolution,grandjean2023ethereum,liu2023demonitor}, or
one month~\cite{valdivia2019decentralization,gochhayat2020measuring,li2021comparison,lin2021measuring,liu2022understanding,liu2023demonitor}. A
single study also used longer windows, including three months, one year, and
even the entire history of the ledger~\cite{gochhayat2020measuring}. Regarding block-based
windows, one study aligned the number of blocks with Bitcoin's difficulty
adjustment period~\cite{romiti2019deep}, while another used a window of 20
blocks~\cite{mueller2022understanding}.

\subsubsection{Population estimation.}
An interesting finding is that estimating the population of block producers was
never identified as a distinct choice in the literature.
Therefore, we assume that the population was inferred from the resource estimation window, which is why the two columns are merged in all entries of
Table \ref{tab:arbitrary-choices}.
As a reminder, the estimation windows are only relevant for the consensus layer,
where the power distribution is not directly available.
In contrast, the wealth distribution in the tokenomics layer can be directly
observed from the blockchain, hence these considerations do not apply (marked
as N/A).

\subsubsection{Measurement frequency.}
In longitudinal studies, the measurement frequency was often conflated with the resource estimation window. 
For instance, some authors used a single term, such as ``granularity'', to refer to both concepts. 
Notably, Lin \etal~\cite{lin2021measuring} were the first to propose the use of sliding windows, effectively decoupling the estimation window from the measurement frequency
(though they still used the term ``granularity'' for both of them).
This approach was also followed by Grandjean \etal~\cite{grandjean2023ethereum},
who employed a ``7-day rolling window'', from which we infer that the measurements were conducted daily, even though not explicitly stated.
On the other hand, several studies analyzed only a single snapshot of the system~\cite{wu2019information,gochhayat2020measuring,li2021comparison,jensen2021decentralized,Jia}.

\subsubsection{Inclusion threshold.}
Approximately half of the reviewed studies employed inclusion thresholds on the
distribution under consideration. 
In the consensus layer, some works quantified decentralization among the top
$10{-}30$ block
producers~\cite{gencer2018decentralization,li2021comparison,liu2022understanding},
while the rest did not apply any threshold. 
In tokenomics, inclusion criteria were more prevalent. Some
researchers focused exclusively on the top-X wealthiest entities, with X ranging
from $30$ to
$10{,}000$~\cite{wu2019information,nadler2020decentralized,kusmierz2022centralized,Jia}.
Others focused on entities that control a minimum amount of tokens, most often above the fees required to issue a
transaction~\cite{jensen2021decentralized,barbereau2022defi}. Nonetheless, the selection of these thresholds is often arbitrary and lacks clear justification.

\subsubsection{Decentralization Metrics.} \label{subsec:metrics}



Several metrics have been employed in the literature to measure blockchain decentralization on the consensus and tokenomics layers. Starting from
concentration ratios~\cite{gencer2018decentralization,valdivia2019decentralization,wang2020measurement,li2021comparison,nadler2020decentralized} --- or ``mining power ratios'' as often referred to in this literature --- various metrics were gradually incorporated into the blockchain decentralization toolkit.
Overall, the Gini coefficient~\cite{romiti2019deep,gochhayat2020measuring,nadler2020decentralized,jensen2021decentralized,lin2021measuring,kusmierz2022centralized,Jia,barbereau2022defi,mueller2022understanding,campajola2023evolution,grandjean2023ethereum,liu2023demonitor,yan2024analyzing} and the Nakamoto coefficient~\cite{gochhayat2020measuring,li2021comparison,nadler2020decentralized,jensen2021decentralized,lin2021measuring,zeng2021characterizing,kusmierz2022centralized,campajola2023evolution,grandjean2023ethereum,liu2023demonitor,yan2024analyzing} were most frequently
used across the reviewed literature.
Notably, there are studies that measure the Nakamoto coefficient without explicitly naming it \cite{romiti2019deep,wang2020measurement}, and the same goes for those that measure its generalization, the $\tau$-decentralization index~\cite{li2021comparison,nadler2020decentralized,grandjean2023ethereum}.
Shannon entropy has also been used extensively over the years~\cite{wu2019information,gochhayat2020measuring,li2021comparison,lin2021measuring,kusmierz2022centralized,Jia,barbereau2022defi,liu2022understanding,liu2023demonitor,yan2024analyzing}, while the Herfindahl–Hirschman index has been included in more recent works~\cite{grandjean2023ethereum,yan2024analyzing}.
In some cases, simply counting the number of participants (block producers or token holders) was also used as a measure of decentralization~\cite{gochhayat2020measuring,nadler2020decentralized,liu2022understanding}.
Notably, no metric was consistently applied across all studies, making it difficult to draw
meaningful comparisons among their results.

\section{Empirical Analysis}\label{sec:empirical}

\begin{figure}[t]
    \centering
    \includegraphics[width=\linewidth]{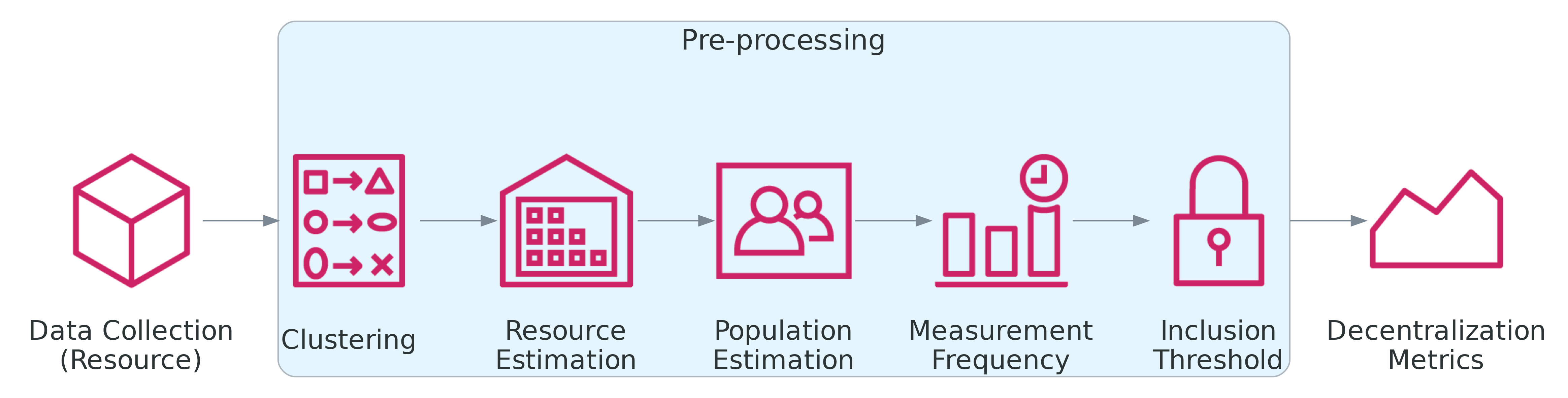}
    \caption{Decentralization measurement pipeline.}
    \label{fig:pipeline}
\end{figure}

We now analyze the impact of the methodological decisions outlined above using data from live
blockchain systems. We collected historical data from five ledgers: Bitcoin, Cardano, Ethereum,
Litecoin, and Tezos. 
Notably, Ethereum transitioned from Proof-of-Work to Proof-of-Stake on September
2022~\cite{kapengut2023event}, so we treat it as two different systems, Ethereum
PoW and Ethereum PoS. 
Our selection is diverse in terms of Sybil resilience
mechanisms\footnote{Bitcoin, Ethereum Pow, and Litecoin are Proof-of-Work.
Cardano, Ethereum PoS, and Tezos are Proof-of-Stake.} and balance
models\footnote{Bitcoin, Cardano, and Litecoin use UTxOs. Ethereum and Tezos are
account-based.}. By analyzing these varied design architectures, we increase confidence that our conclusions generalize, especially when results are consistent
across systems. 
In the paragraphs below, we discuss our approach 
for each step of the process.
Figure~\ref{fig:pipeline} also depicts the pipeline of our analysis.

\subsubsection{Data Collection (Resource).}
In line with most prior work, we used blocks as the resource for estimating power
distributions on the consensus layer. 
For the tokenomics layer, we use the system's native token as the resource, an
approach consistently adopted in the relevant literature.
As discussed in Section~\ref{sec:framework}, we do not analyze the governance layer due to lack of sufficient measurement studies on layer-1 blockchains.
We collected historical data for both layers from Big Query's public
datasets~\cite{fernandes2015bigquery} and, in some cases, from our own full nodes.
Due to page limit constraints, we refer to Appendix~\ref{app:data} for more information on the collection process and a description of the relevant data. 


\subsubsection{Clustering.}
For both layers, we collected attribution data to determine how resources are distributed among real-world entities rather than 
addresses. 
This includes tags associated with known pools, addresses controlled by known entities (\eg exchanges), and information on legal links between organizations, such as acquisitions of mining pools.\footnote{For example, the mining pool BTC.COM was
acquired BIT Mining~\cite{btcm_press_316}, so after
the point of the acquisition, blocks created by BTC.COM are attributed to BIT
Mining.} 
We use blockchain explorers\footnote{Such explorers are \href{https://etherscan.io}{Etherscan}, \href{https://bitinfocharts.com/}{BitInfoCharts}, and
\href{https://www.blockchain.com}{blockchain.com}.} and community
projects~\cite{bitcoinDataMiningPools} to collect off-chain data.
For Bitcoin, we also implemented the multi-input heuristic~\cite{DBLP:conf/socialcom/ReidH11}, using the transaction graph retrieved from a full node, 
and for Cardano, we grouped addresses that shared the same stake key~\cite{chegenizadeh2023heuristic}.

\begin{figure}[t]
    \centering
    \includegraphics[width=0.8\linewidth]{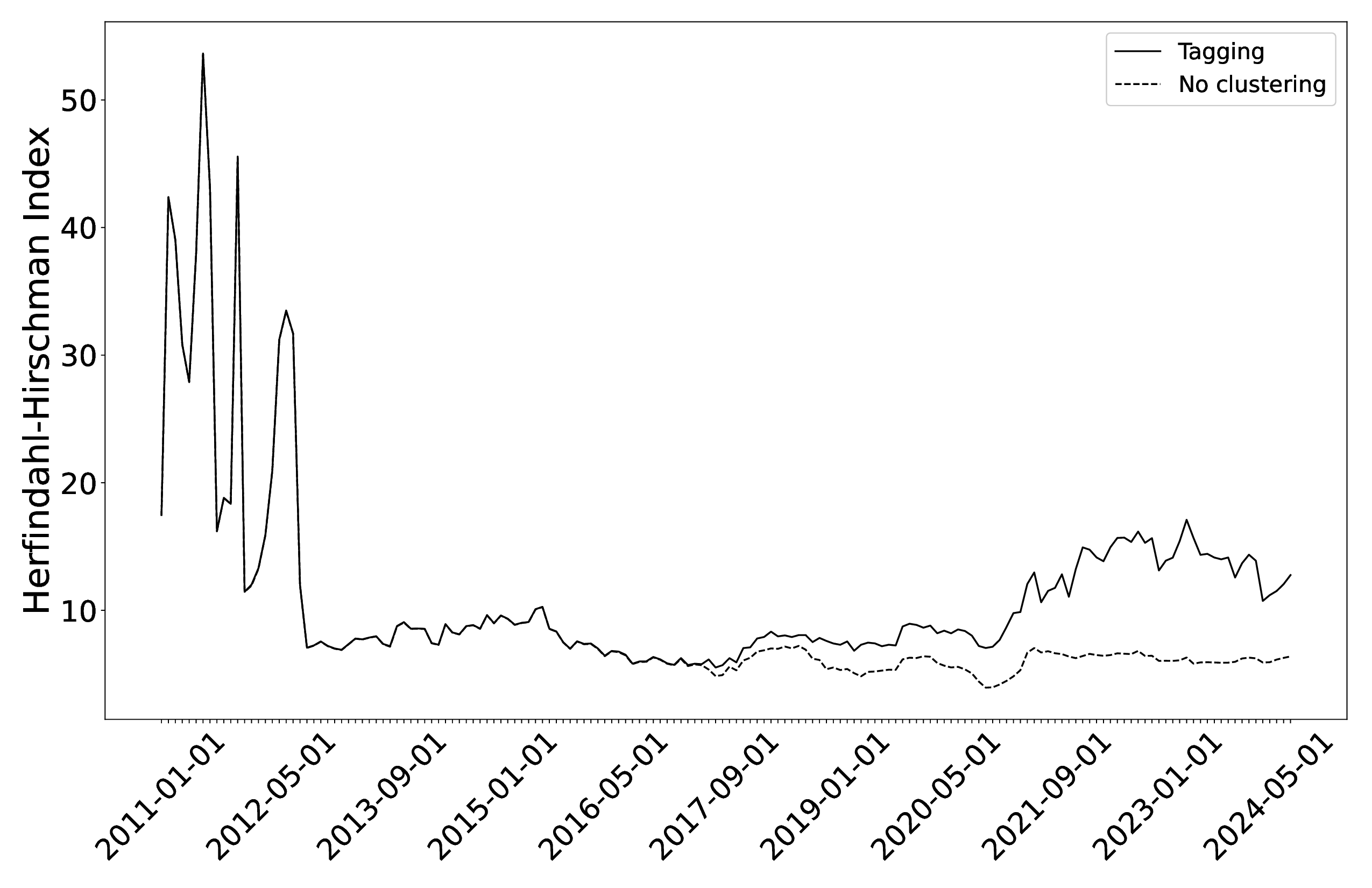}
    \caption{Bitcoin tokenomics HHI values with and without clustering (tagging). }
    \label{fig:clustering-comparison}
\end{figure}

We conduct an analysis to assess the effectiveness of tagging, as this has not been thoroughly addressed in the literature. 
Figure \ref{fig:clustering-comparison} illustrates the HHI results for Bitcoin on the tokenomics layer, comparing the outcomes of tagging as a clustering method (solid lines) with the baseline of unclustered data (dashed lines).
From 2016 onward, the two lines begin to diverge, indicating that tagging successfully grouped enough addresses to influence the results.\footnote{Recall that higher HHI signifies greater centralization.}
As expected, clustering increases centralization under the HHI, since it aggregates the same resources into fewer entities.\footnote{Notably, this would not necessarily be true under the Gini Coefficient.}
\looseness=-1

\subsubsection{Resource Estimation Window.}
Prior research on estimation windows for the consensus layer suggests that short windows, including the 1-day window used occasionally in the literature, do not provide enough statistical confidence to reliably estimate metrics like the Nakamoto Coefficient, especially for low throughput ledgers like Bitcoin~\cite{milad2024statistical}.
In contrast, windows of 7 or more days were shown to provide statistical confidence in the results.
Therefore, we adopted a 7-day window for aggregating blocks and estimating the distribution of block production power.
\looseness=-1 



\subsubsection{Population Estimation.}

\begin{figure}[t]
    \centering
    \includegraphics[width=0.8\linewidth]{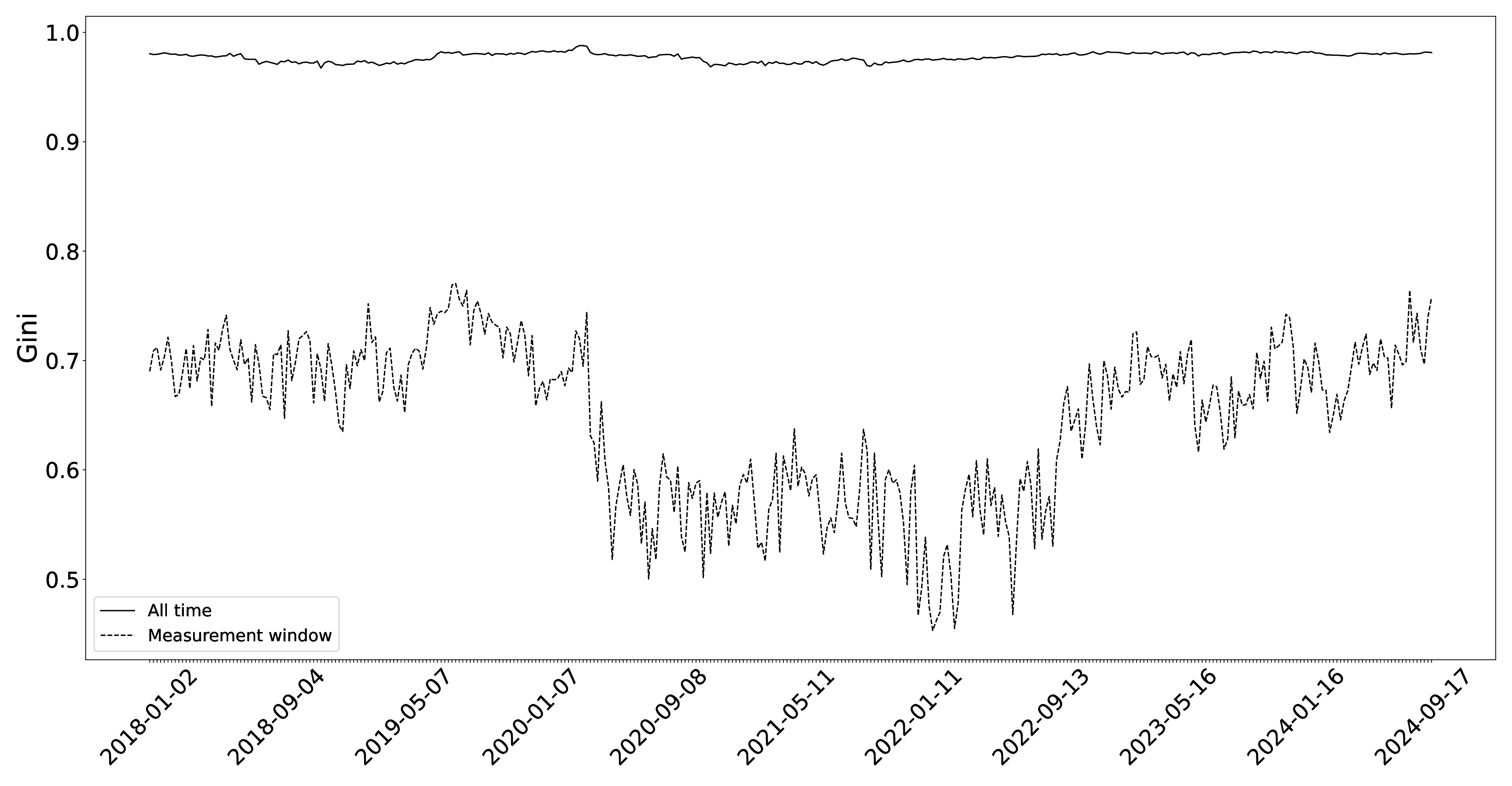}
    \caption{Historical Gini coefficient of Bitcoin using different population estimation windows. `All Time' assumes an entity who mined once is active for the entire study window, whereas `Measurement Window' ignores miners who were unsuccessful in a given estimation window.}
    \label{fig:btc-gini-population-windows}
\end{figure}

We opted for the simple method of using an additional observation window for estimating the population of block producers at a given point in time.
We evaluated the effects of different window sizes, to determine how the population estimation window can impact decentralization estimates.
Figure \ref{fig:btc-gini-population-windows} shows the results of
the Gini coefficient on weekly Bitcoin data, using two different
approaches for estimating the population of block producers. For the solid line
the population includes all entities that created at least one block
between $2018{-}2024$.
This estimate includes miners who were inactive during certain periods.
In contrast, the dashed line only
accounts for the entities that produced blocks in each specific week.
This estimate ignores active participants that were unsuccessful in producing blocks during the resource estimation window.
The significant discrepancy between these two values shows that the population window is an important criterion when estimating decentralization.
In our analysis, we use the entire history for population estimation, leaving a thorough evaluation of the optimal choice to future work.

\subsubsection{Measurement Frequency.}
In a longitudinal study, the measurement frequency, together with the size of
the study window, determines the number of data points for analysis. 
Although measurement frequency is a distinct choice, it should be selected in
conjunction with estimation windows, as it determines whether the data contains
overlapping or non-overlapping windows. 
For example, if the estimation window is seven days, weekly sampling produces
non-overlapping windows, whereas daily sampling results in overlapping windows.

Overlapping windows are appropriate for descriptive statistics.
However, when conducting inferential statistics, overlapping windows inflate the
number of data points without meaningfully increasing the number of observations,
since the same underlying data is repeatedly counted.
As our analysis involves such statistics (see metrics paragraph below), we opted for non-overlapping windows.
Specifically, we sample consensus data weekly and tokenomics data monthly.

%

\subsubsection{Inclusion Thresholds.}

We also explored the role of thresholds, given they were widely used in prior
work, particularly for the tokenomics layer. 
The motivation behind using thresholds is to provide clearer insights, by
removing low-resource entities, such as unusable ``dust'' wallets. However, it
is unclear whether the results of decentralization analyses are robust to
different threshold choices. 

We tested all types of thresholds found the literature, \ie accounting only for the top X wealthiest entities (with X ranging from 100 to 10,000), entities that control enough tokens to issue transactions, or all entities. 
We also included an additional type of threshold that considers only the top X\% wealthiest entities --- specifically we looked at the top 50\%.  
Table~\ref{tab:btc-gini-thresholds} presents the Spearman correlations between the Gini values for Bitcoin, calculated on tokenomics data using the various inclusion thresholds.

\begin{table}[t]
    \centering
\begin{tabular}{|l|rrrrrrr|}
 \hline
 & Top 100 & Top 500 & Top 1000 & Top 10000 & Top 50\% & > tx fees &  All\\
 \hline
Top 100 & {\cellcolor{green}} 1.00 & {\cellcolor{yellow}} 0.17 & {\cellcolor{yellow}} -0.17 & {\cellcolor{GreenYellow}} 0.30 & {\cellcolor{yellow}} 0.14 & {\cellcolor{yellow}} 0.16 & {\cellcolor{yellow}} 0.16 \\
Top 500 & {\cellcolor{yellow}} 0.17 & {\cellcolor{green}} 1.00 & {\cellcolor{LimeGreen}} 0.82 & {\cellcolor{GreenYellow}} 0.34 & {\cellcolor{pink}} -0.49 & {\cellcolor{OrangeRed}} -0.71 & {\cellcolor{yellow}} -0.20 \\
Top 1000 & {\cellcolor{yellow}} -0.17 & {\cellcolor{LimeGreen}} 0.82 & {\cellcolor{green}} 1.00 & {\cellcolor{yellow}} 0.24 & {\cellcolor{OrangeRed}} -0.78 & {\cellcolor{red}} -0.93 & {\cellcolor{red}} -0.93 \\
Top 10000 & {\cellcolor{GreenYellow}} 0.30 & {\cellcolor{GreenYellow}} 0.34 & {\cellcolor{yellow}} 0.24 & {\cellcolor{green}} 1.00 & {\cellcolor{orange}} -0.54 & {\cellcolor{yellow}} -0.29 & {\cellcolor{OrangeRed}} -0.76 \\
Top 50\% & {\cellcolor{yellow}} 0.14 & {\cellcolor{pink}} -0.49 & {\cellcolor{OrangeRed}} -0.78 & {\cellcolor{orange}} -0.54 & {\cellcolor{green}} 1.00 & {\cellcolor{LimeGreen}} 0.88 & {\cellcolor{LimeGreen}} 0.88 \\
> tx fees & {\cellcolor{yellow}} 0.16 & {\cellcolor{OrangeRed}} -0.71 & {\cellcolor{red}} -0.93 & {\cellcolor{yellow}} -0.29 & {\cellcolor{LimeGreen}} 0.88 & {\cellcolor{green}} 1.00 & {\cellcolor{green}} 1.00 \\
 & {\cellcolor{yellow}} 0.16 & {\cellcolor{yellow}} -0.20 & {\cellcolor{red}} -0.93 & {\cellcolor{OrangeRed}} -0.76 & {\cellcolor{LimeGreen}} 0.88 & {\cellcolor{green}} 1.00 & {\cellcolor{green}} 1.00 \\
  \hline
\end{tabular}

    \caption{Spearman correlations of Bitcoin's historical Gini coefficients on tokenomics data, calculated with different inclusion thresholds.}
    \label{tab:btc-gini-thresholds}
\end{table}

Our analysis suggests that thresholds focused on the top X entities show little correlation with one another.
This means that the arbitrary decision of which threshold to choose (Top 100 vs 500 vs 1000) meaningfully changes estimates of decentralization trends.
In contrast, the remaining thresholds (top 50\%, more than the transaction fees, and all entities) show high correlation.
Note, this does not mean that these three thresholds result in the same values in absolute terms, but rather that they capture the same trends.
In the rest of our experiments, we continue by using the entire distributions, without applying any thresholds. 

\subsubsection{Decentralization Metrics.}
\label{subsec:empirical-metrics}
Our objective in this part of the analysis is to examine the relationships between the various metrics used in the literature and determine whether they capture a common underlying notion of decentralization.
To achieve this, we employ exploratory factor
analysis (EFA)~\cite{fabrigar2011exploratory}, a statistical tool that has been widely used across different disciplines, including in studies of government decentralization~\cite{schneider2003decentralization}. 

In EFA, the observed variables --- in this case the metrics --- are analyzed to infer latent variables, or factors, that capture the data in a more succinct manner. 
If all metrics perfectly captured the same underlying notion of decentralization, the analysis would reveal a single factor.
Conversely, if the metrics reflect different constructs, the analysis would produce multiple factors.
In such cases, we expect metrics representing similar aspects to group together by loading onto the same factor. 
We also analyze correlations between metrics to confirm the results in a more intuitive way.
The detailed methodology we followed, including adequacy tests and data transformations, is provided in Appendix~\ref{app:efa}.

Figure \ref{fig:EFA-consensus} illustrates the results of the factor analysis on the consensus layer, with the metrics calculated using 7-day resource estimation windows, all-time population estimation window, 7-day frequency, clustering (tagging) and no thresholds.
We observe that for three of the ledgers (Cardano, Ethereum PoS and Tezos) all metrics load onto a single factor, suggesting they capture the same underlying concept of decentralization.
However, the other three (Bitcoin, Ethereum PoW and Litecoin) all exhibit two factors, with the number of parties consistently loading strongly onto the second factor, suggesting that it captures a different signal compared to the other metrics.
The pairwise Spearman correlations between the metrics (see Table~\ref{tab:consensus-correlations-all} in the Appendix) also reveal that the number of participants in these blockchains is either uncorrelated or even inversely correlated with other decentralization metrics.

\begin{figure*}[t]
    \centering
    \includegraphics[width=\textwidth]{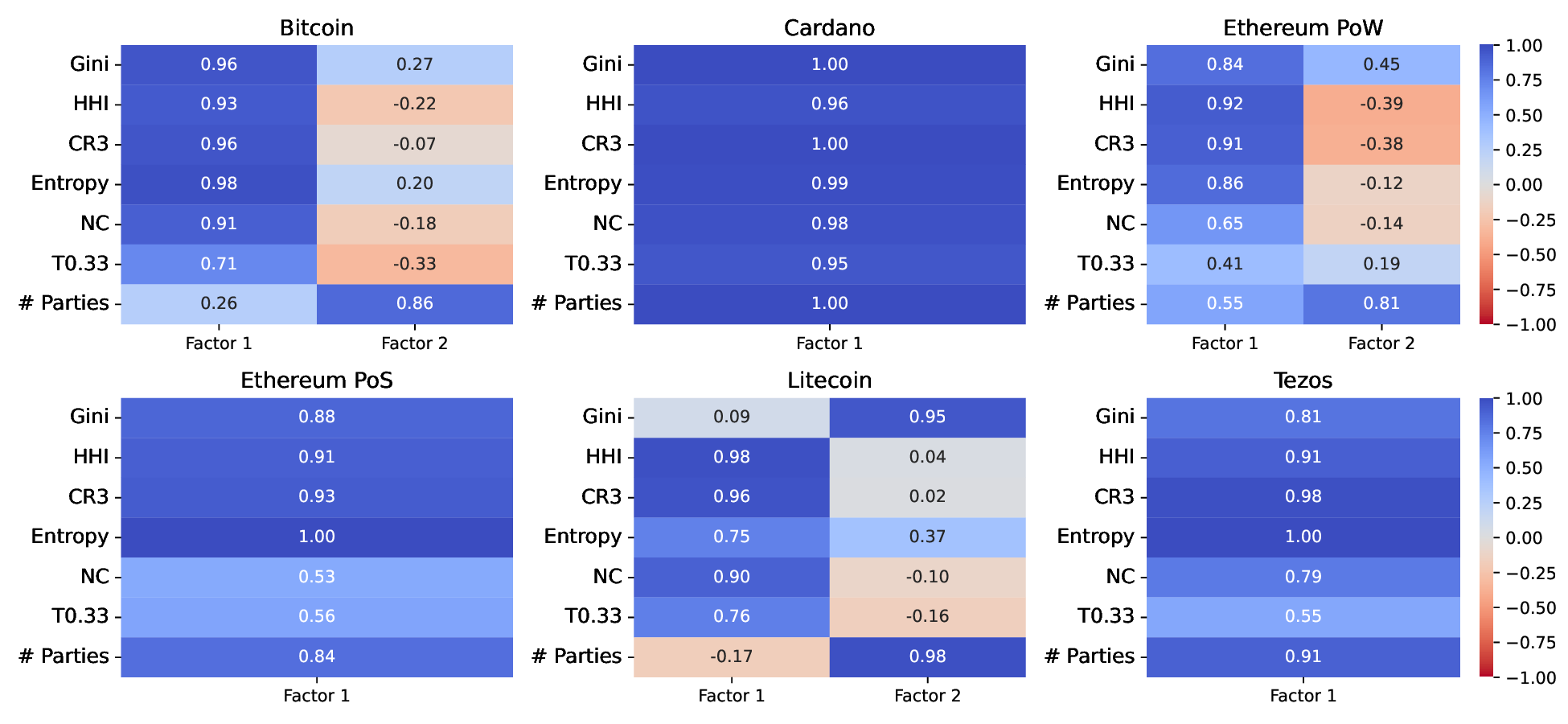}
    \caption{Factor analysis results for decentralization metrics on the consensus layer.}
    \label{fig:EFA-consensus}
\end{figure*}

The same analysis was conducted for tokenomics, with the metrics calculated using monthly frequency, clustering (tagging) and no thresholds (see Figure~\ref{fig:EFA-tokenomics} in the Appendix).
However, the results were more nuanced on this layer, with no consistent narrative emerging across the different systems. 

\section{Discussion and Recommendations}
\label{sec:recommendations}
This section discusses our systematization and measurement results, with a focus on offering forward-looking recommendations.
These recommendations are derived through reasoning from first principles and, where applicable, by interpreting the findings of our empirical analysis.
Table~\ref{tab:recommendations} summarizes our suggestions for which options to choose and which to avoid when addressing the various dimensions of the decentralization measurement process. 
Additionally, the table connects our recommendations to the reviewed literature, offering an overview of the proportion of studies that adopted each choice.

\subsubsection{Resource.}
The choice of resource is straightforward in tokenomics, as there is general agreement that tokens serve as the primary resource for this layer. 
In contrast, the consensus layer offers two approaches commonly used in the literature. 
We propose using blocks as the resource, as they better reflect participation in the consensus process. 
The alternative method, which relies on block rewards, is less robust due to variations in reward structures across different systems, and even within a single system over time.

\subsubsection{Clustering.}
Figure~\ref{fig:clustering-comparison} demonstrated that tagging had a significant influence on Bitcoin's HHI estimates since 2016.
The lack of difference before then is likely because the information sources did not tag legacy blocks.
Based on this observation, we conclude that future studies can reasonably rely on tagging, particularly when analyzing established blockchains with sufficient public attribution data.
In addition, implementing heuristics tailored to the architecture of a blockchain system --- such as the multi-input heuristic for UTXO-based ledgers --- can further enhance the clustering results, as demonstrated in the literature for Bitcoin~\cite{DBLP:journals/cacm/MeiklejohnPJLMV16,DBLP:conf/uic/HarriganF16}, Ethereum~\cite{victor2020address} and Cardano~\cite{chegenizadeh2023heuristic}.

\subsubsection{Resource estimation window.}
As discussed earlier, prior work has already addressed the question of choosing an appropriate window size for estimating resource distribution in blockchains.
In line with~\cite{milad2024statistical}, we recommend using a window of at least 7 days when estimating block distributions, especially for low throughput ledgers like Bitcoin.
Additionally, we caution researchers against using windows smaller than 150 blocks, as it has been demonstrated that Bitcoin failed the relevant hypothesis test in over half of the cases when aggregating over a daily window, which roughly corresponds to this block count~\cite{milad2024statistical}.

\subsubsection{Population estimation.}
Figure~\ref{fig:btc-gini-population-windows} reveals how influential the size of the observation window can be when estimating the population at a given time period.
The population window must strike a balance between capturing the unsuccessful but active participants, while also excluding those inactive during the relevant time
period.
We recommend choosing a window that is larger than the resource estimation window but smaller than the study window (\ie the total observation period).
A rule of thumb could be to use a population window twice the size of the resource window. 
Further research is needed to determine whether more sophisticated methods, such as statistical inference, could provide more accurate population estimates compared to this heuristic.

\subsubsection{Measurement frequency.}
The data in a longitudinal analysis should be sampled frequently enough to capture temporal changes.
For example, sampling the data only once a year might overlook significant trends that occurred throughout the year. 
We recommend experimenting with various sampling intervals to ensure the chosen frequency does not obscure significant change between measurements. 
We advise against using overlapping windows --- \ie  sampling more frequently than the resource estimation window --- when performing inferential statistics, as this can inflate the number of data points without adding observations.

\subsubsection{Inclusion threshold.}
At the consensus layer, an entity that produces a block is, by definition, active.
Consequently, applying thresholds there would inevitably exclude active participants, leading to skewed decentralization estimates. 
For this reason, we advise against the use of thresholds in the consensus layer.

In contrast, thresholds may be justified in the tokenomics layer, as inactive addresses can still hold tokens.
However, Table~\ref{tab:btc-gini-thresholds} demonstrated that the different top-X thresholds used in the literature yield decentralization estimates that are not correlated with each other.
This indicates that arbitrary threshold choices can significantly impact results, potentially enabling manipulation by selectively adjusting the threshold.
Interestingly, other types of thresholds showed stronger correlations with one another and with estimates derived from the entire distribution, suggesting that they provide more robust and reliable results. 
We consider the transaction fee threshold to be a reasonable choice, as it effectively excludes entities with no economic power in the system, while ensuring that the broader distribution remains representative.

\begin{table}[t]
\scriptsize
\begin{tabular}{|c|cc|cc|}
\hline
\multirow{2}{*}{\textbf{Workflow step}} & \multicolumn{2}{c|}{\textbf{Recommended}} & \multicolumn{2}{c|}{\textbf{Discouraged}} \\ \cline{2-5} 
 & \multicolumn{1}{c|}{\textit{Decision}} & \textit{Adoption} & \multicolumn{1}{c|}{\textit{Decision}} & \textit{Adoption} \\ \hline
Resource & \multicolumn{1}{c|}{\begin{tabular}[c]{@{}c@{}}Blocks (consensus)\\ Tokens (tokenomics)\end{tabular}} & \begin{tabular}[c]{@{}c@{}}13/15\\ 7/7\end{tabular} & \multicolumn{1}{c|}{Rewards (consensus)} & 2/15 \\ \hline
Clustering & \multicolumn{1}{c|}{\begin{tabular}[c]{@{}c@{}}Tagging\\ Heuristics\end{tabular}} & \begin{tabular}[c]{@{}c@{}}5/20\\ 5/20\end{tabular} & \multicolumn{1}{c|}{No clustering} & 14/20 \\ \hline
\begin{tabular}[c]{@{}c@{}}Resource estimation\\ window (REW)\end{tabular} & \multicolumn{1}{c|}{$\geq$ 7 days} & 12/15 & \multicolumn{1}{c|}{< 150 blocks} & 5/15 \\ \hline
\begin{tabular}[c]{@{}c@{}}Population estimation\\ window (PEW)\end{tabular} & \multicolumn{1}{c|}{REW <  PEW < study window} & 0/15 & \multicolumn{1}{c|}{PEW < REW} & 0/15 \\ \hline
\begin{tabular}[c]{@{}c@{}}Inclusion\\ threshold\end{tabular} & \multicolumn{1}{c|}{\begin{tabular}[c]{@{}c@{}}None (consensus)\\ > tx fees (tokenomics)\end{tabular}} & \begin{tabular}[c]{@{}c@{}}12/15\\ 2/7\end{tabular} & \multicolumn{1}{c|}{Top X} & 7/20 \\ \hline
\begin{tabular}[c]{@{}c@{}}Decentralization\\ metrics\end{tabular} & \multicolumn{1}{c|}{NC, HHI (consensus)} & 9/15 & \multicolumn{1}{c|}{\# Parties} & 3/20 \\ \hline
\end{tabular}
\caption{Summary of our recommendations for various steps of the decentralization measurement workflow, along with the adoption rates for each recommended or discouraged choice based on the relevant articles listed in Table \ref{tab:arbitrary-choices}. Note that the denominators in the adoption rates vary, as some decisions are only applicable to specific contexts (\eg the consensus layer) and therefore only pertain to studies addressing those cases.}
\label{tab:recommendations}
\end{table} 

\subsubsection{Metrics}

The results of the factor analysis on the consensus layer (Figure~\ref{fig:EFA-consensus}) suggest that the blockchains we examined are divided into two categories.
In the first category, all metrics load onto a single factor, whereas in the second two factors emerge, with the number of parties consistently loading onto the second factor, separate from most other metrics.
Upon closer inspection, we observe that the first group comprises only PoS blockchains (Cardano, Ethereum PoS and Tezos), while the second only PoW ones (Bitcoin, Ethereum PoW and Litecoin).
This allows us to speculate that this architectural choice has an impact on how effectively a blockchain system can translate participation into decentralization.

In particular, in PoW systems, decentralization and participation emerge as distinct factors, suggesting that increasing participation is not sufficient to achieve greater decentralization. 
This may be due to economies of scale, where larger mining operations benefit from lower costs and more efficient hardware, thereby consolidating control over block production.
At any rate, it appears that merely counting the number of participants is not representative of decentralization, so we advise against using it in decentralization analyses of these systems. 

In contrast, for PoS blockchains, all metrics cluster onto a single factor, which we interpret as decentralization itself, as the metrics were selected to capture this concept. 
The fact that the number of parties is clustered together with the other metrics here indicates that PoS blockchains are more effective at aligning decentralization with participation than their PoW counterparts.


Having identified that many of the metrics used in prior work capture the same underlying signal, the next goal is to decide on the most useful ones.
This can only be done by reasoning from first principles, as there is no ground truth with which to validate metrics.
We proceed by discussing each metric in turn.

Entropy and Gini are both tail-sensitive metrics that take the entire distribution into consideration.
The main strengths of entropy are its simple mathematical definition and long scientific heritage.
However, it has weaknesses when it is applied to quantifying blockchain decentralization.
Most fundamentally, a blockchain for which one actor holds more than 50\% of the resources can have an arbitrarily high entropy value if the number of participants is sufficiently high. 
This means relying on entropy to measure decentralization could create a false sense of security.
Another problem lies in interpreting values as there is no clear rationale, unlike some other metrics. 

Gini was designed to quantify how fairly resources, be it wealth or income, are distributed across society.
This allows for comparisons between centralization of tokens, a form of wealth, and centralization of wealth in society, which can help with interpreting values.
Gini also has an intuitive graphical explanation. 
However, it is highly sensitive to the tail of the distribution, which can pose challenges in blockchain measurements if the clustering of addresses is not entirely successful. 
Inaccurate clustering may lead to skewed Gini values, potentially misrepresenting the level of decentralization.

In contrast, the Nakamoto coefficient, $\mathsf{\tau}$-decentralization index and concentrations ratio are all tail-insensitive metrics that only account for the head of a distribution.
%
The Nakamoto Coefficient and $\tau$ are better in that they capture information about how many parties need to coordinate to subvert the system, when the $\tau$ threshold is set with specific attacks in mind.
While $\tau = 0.5$, aka the Nakamoto Coefficient, can warn about the potential for 51\% attacks, it is not designed to warn about similar attacks in systems with security thresholds lower than $50\%$, such as Algorand~\cite{gilad2017algorand} or Ethereum PoS~\cite{buterin2020combining,buterin2020incentives}, which require two thirds of consensus participants to be honest.
We suggest using the Nakamoto coefficient when measuring consensus decentralization of systems that require an honest majority, and $\tau=0.33$ for systems that require $\frac{2}{3}$ of honest participants.

Finally, HHI includes a contribution from the tail of the distribution, but discounts it via a quadratic term. 
Originally designed to quantify market concentration, the HHI is easily adaptable to the block production market of blockchains. Its values are also easier to interpret thanks to established guidelines~\cite{united2010horizontal}.




\section{Conclusion}
Our work systematizes the different ways of measuring 
decentralization in blockchain systems.
We developed a framework that distills the decentralization measurement process into three distinct steps:
(i) determining which data to collect (layer / resource),
(ii) extracting the distribution of resources,
and 
(iii) choosing which decentralization metrics to use. 
Notably, step (ii) requires a number of impactful choices
to be made in a form of pre-processing, which involve 
(a) clustering,
(b) resource estimation,
(c) population estimation,
(d) measurement frequency, 
and (e) inclusion thresholds.
%
We reviewed existing literature and classified it based on our systematization framework.
Our investigation revealed that prior work lacks a systematic way for choosing which metrics to use and which pre-processing steps to follow --- instead, most such choices were made in an arbitrary manner.
%


To make the landscape of various choices easier to navigate, we conduct an empirical analysis using data from five different blockchains on the consensus and tokenomics layers. 
We combine the results of the analysis with  first-principles reasoning to derive recommendations for future researchers.
Our findings advise against some of the approaches that have been employed in the literature, such as examining only the top X wealthiest entities when measuring decentralization on the tokenomics layer.
We also show that the choice of population estimation window, which has not received any attention in prior work, can significantly impact decentralization estimates when using metrics that take the entire population into account, such as the Gini coefficient.

A large part of our analysis was dedicated to exploring the relationship between the different metrics that have been proposed for measuring decentralization, 
and using factor analysis to determine whether they capture the same latent constructs.
%
%
%
An interesting finding that stemmed from this analysis is that in Proof-of-Work blockchains participation on the consensus layer is not correlated with decentralization, but rather captures a distinct signal.
This suggests that, perhaps contrary to popular belief, higher levels of participation are not always associated with greater decentralization.
%

This insight has significant implications for the design and governance of blockchain systems. Developers and stakeholders should consider mechanisms that not only encourage broad participation but also ensure that control and decision-making processes are sufficiently distributed. 



\subsubsection{Future work.}
Our work motivates various avenues for further research.
While we are confident in our recommendations at the consensus layer, tokenomics presents additional challenges.
Token clustering is more difficult because participants do not disclose token ownership, unlike mining pools that typically advertise which blocks they produced.
Further, it is hard to know whether an address is abandoned forever or simply storing cryptocurrency long-term.
    Additionally, identifying the optimal mechanism for population estimation is a challenging, albeit necessary, step.
The community should prioritize research that addresses these problems.
\looseness=-1


Improving decentralization measurement for other layers of blockchain systems, such as network or governance, will involve solving two core problems.
First, researchers need to address inconsistent and missing data when relying on off-chain data sources.
Second, fungible resources must be identified for these layers, or methods must be developed to appropriately weigh non-fungible resources 
according to their influence.
We recommend that researchers conduct primary research in this area before attempting premature systematization.

Finally, the community adopting standardized measures of decentralization could unlock exciting research opportunities.
For example, event studies could analyze how decentralization dynamics shift in response to governance decisions, regulatory changes, or market shocks.
This line of research could help identify which specific mechanisms or components  (\eg more sophisticated reward sharing schemes or proposer-builder separation methods)
are successful in translating participation into decentralization in the context of a blockchain system, which is particularly relevant given the inconsistencies observed in PoW blockchains.

\medskip

%
%
%
\bibliographystyle{splncs04}
\bibliography{references.bib}

\appendix
\section{Appendix}


\subsection{Data Description and Sources}
\label{app:data}
For the consensus layer, we collect data about the blocks that were produced in each ledger. Specifically, for each block, we obtain its number, timestamp, address(es) that received rewards for it and any ledger-specific fields that can potentially be used to identify its creator (for example, in Bitcoin this can be an encoded ``tag'' of a mining pool).
For tokenomics, we collect data about each address that holds a non-zero amount of the system's native token and its corresponding ``balance''. Note that in account-based systems, such as Ethereum, the balance of an address can be retrieved directly from the ledger, while in UTXO-based systems, such as Bitcoin, the balance is derived from all the inputs and outputs of the transactions that an address has been involved in.
Both block and balance information, as described above, can be obtained directly from the blockchain. 
We collect these on-chain data from Google Big Query's public
datasets~\cite{fernandes2015bigquery}. 

\begin{table}[t]
    \centering
    \begin{tabular}{|c|c|c|}
    \hline
        \textbf{Ledger} & \textbf{Observation period} & \textbf{\#Blocks} \\
        \hline
         Bitcoin & Jan 2018 - Oct 2024 & 361,762 \\
         Cardano & Jan 2018 - Dec 2023 & 6,858,920 \\
         Ethereum PoW & Jan 2018 - Sep 2022 & 10,709,818 \\
         Ethereum PoS &  Sep 2022 - Oct 2024 & 5,331,582 \\
         Litecoin & Jan 2018 - Sep 2024 & 1,423,055 \\
         Tezos & Jun 2018 - Dec 2022 & 2,926,050 \\
        \hline
    \end{tabular}
    \caption{Observation period and number of blocks captured for each ledger on the consensus layer.}
    \label{tab:data_consensus}
\end{table}

\begin{table}[t]
    \centering
    \begin{tabular}{|c|c|}
    \hline
        \textbf{Ledger} & \textbf{Observation period} \\ 
        \hline
         Bitcoin & Jan 2011 - Aug 2024\\ 
         Cardano & Jan 2018 - Dec 2024\\ 
         Ethereum PoW & Jan 2016 - Sep 2022\\ 
         Ethereum PoS & Sep 2022 - Aug 2023\\ 
         Litecoin & Jan 2013 - Aug 2024\\ 
         Tezos & Jan 2019 - Aug 2024\\ 
        \hline
    \end{tabular}
    \caption{Observation period for each ledger on the tokenomics layer.}
    \label{tab:data_tokenomics}
\end{table}

The data we collected for each ledger spans at least 5 years 
, with monthly observations.
A special case is Ethereum, where we have (almost) 5 years of data for Ethereum PoW, but only one year of data for  Ethereum PoS, starting from when the ``merge'' took place in 2022.
Tables \ref{tab:data_consensus} and \ref{tab:data_tokenomics} provide details on the time period that was used for the different blockchains for each layer, as well as some additional information, such as the number of blocks captured in the data.



\subsection{Exploratory Factor Analysis}
\label{app:efa}
As exploratory factor analysis is sensitive to outliers~\cite{fabrigar2011exploratory}, we first identify and treat outliers, defined as values that are greater than 3 standard deviations away from the mean.
For treatment, we attempted several popular transformations and observed their effect on the data, in line with prior work~\cite{fabrigar2011exploratory}.
We then used the Kaiser-Meyer-Olkin (KMO) test to determine whether our data was adequate for factor analysis. The KMO value is calculated with the following formula: 
$\text{KMO} = \frac{\sum_{i\neq j}R_{ij}^2}{\sum_{i\neq j}R_{ij}^2 + \sum_{i\neq j}U_{ij}^2}$, 
where $R_{ij}$ is the correlation matrix and $U_{ij}$ is the partial covariance matrix~\cite{factoranalysis}. Possible values range from 0 to 1, with values $>0.5$ considered suitable for factor analysis. All three versions of our data (untouched, outliers removed, and outliers transformed) were deemed suitable across all ledgers.
The Box-Cox transformation was found to most uniformly reduce the number of outliers while still passing relevant adequacy tests, so we adopted that one for the analysis. 

We then determine the appropriate number of factors to be extracted from the data, using eigenvalues according to the Kaiser criterion, which  states that eigenvalues greater than 1 are considered to be significant~\cite{fabrigar2011exploratory}. 
Then, we calculate the loadings using the number of latent factors suggested by the eigenvalues. We also tested a number of orthogonal and oblique rotations to best capture these factors, but opted for the default (promax) because it made little difference to the results.

The factor loadings inform us of which latent variables describe which metrics, and therefore, which metrics tend to clump together. 
The higher the absolute value of a loading, the bigger the influence that the factor has on the metric.
If two metrics load onto the same factor, but with different signs (positive / negative), it means that they capture the same concept, but in different directions. 

\subsubsection{Tokenomics EFA results}
\begin{figure*}[t]
    \centering
    \includegraphics[width=\textwidth]{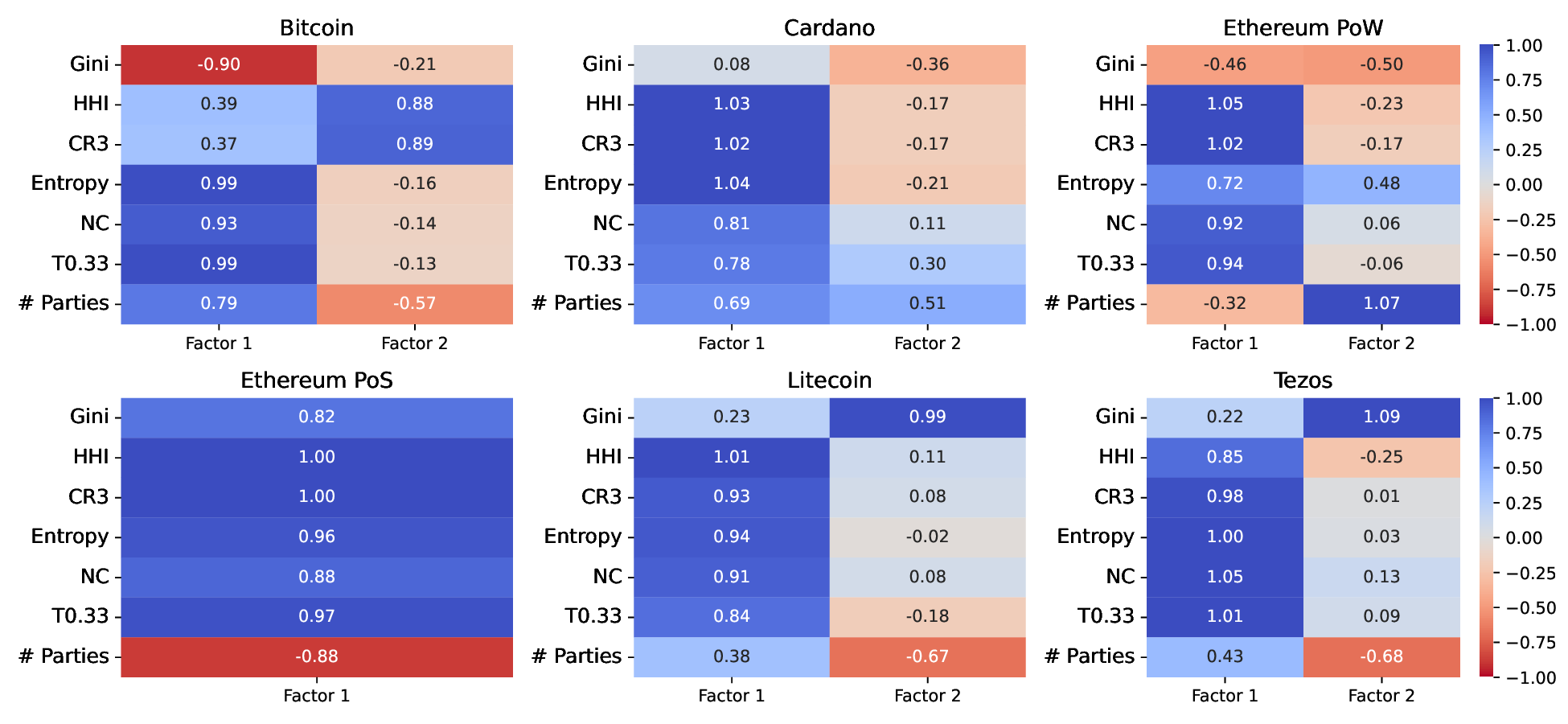}
    \caption{Factor analysis results for decentralization metrics on the tokenomics layer.}
    \label{fig:EFA-tokenomics}
\end{figure*}

Figure~\ref{fig:EFA-tokenomics} presents the results of the exploratory factor analysis on the tokenomic layer.
The metrics were calculated using monthly sampling, clustering (tagging) and no thresholds. 
We observe a more nuanced picture here, making it harder to find common patterns across ledgers.


\subsection{Correlation tables}
Tables \ref{tab:tokenomics-correlations-all} 
and \ref{tab:btc-gini-thresholds} present the Spearman correlations between the various decentralization metrics on the consensus and tokenomics layers.. 








\begin{table*}[t]
\centering

\vspace{0.05in}
\begin{tabular}{|l|rrrrrrr|}
\multicolumn{8}{c}{\small{Bitcoin}} \\
\hline
 & Gini & HHI & CR3 & Entropy & NC & T0.33 & \# Parties \\
 \hline
Gini & {\cellcolor{green}} 1.00 & {\cellcolor{LimeGreen}} 0.82 & {\cellcolor{green}} 0.93 & {\cellcolor{green}} 0.96 & {\cellcolor{LimeGreen}} 0.82 & {\cellcolor{lime}} 0.51 & {\cellcolor{GreenYellow}} 0.36 \\
HHI & {\cellcolor{LimeGreen}} 0.82 & {\cellcolor{green}} 1.00 & {\cellcolor{green}} 0.91 & {\cellcolor{green}} 0.91 & {\cellcolor{green}} 0.93 & {\cellcolor{LimeGreen}} 0.83 & {\cellcolor{yellow}} -0.02 \\
CR3 & {\cellcolor{green}} 0.93 & {\cellcolor{green}} 0.91 & {\cellcolor{green}} 1.00 & {\cellcolor{green}} 0.92 & {\cellcolor{green}} 0.92 & {\cellcolor{lime}} 0.63 & {\cellcolor{yellow}} 0.10 \\
Entropy & {\cellcolor{green}} 0.96 & {\cellcolor{green}} 0.91 & {\cellcolor{green}} 0.92 & {\cellcolor{green}} 1.00 & {\cellcolor{LimeGreen}} 0.86 & {\cellcolor{lime}} 0.66 & {\cellcolor{yellow}} 0.27 \\
NC & {\cellcolor{LimeGreen}} 0.82 & {\cellcolor{green}} 0.93 & {\cellcolor{green}} 0.92 & {\cellcolor{LimeGreen}} 0.86 & {\cellcolor{green}} 1.00 & {\cellcolor{lime}} 0.69 & {\cellcolor{yellow}} -0.04 \\
T0.33 & {\cellcolor{lime}} 0.51 & {\cellcolor{LimeGreen}} 0.83 & {\cellcolor{lime}} 0.63 & {\cellcolor{lime}} 0.66 & {\cellcolor{lime}} 0.69 & {\cellcolor{green}} 1.00 & {\cellcolor{yellow}} -0.07 \\
\# Parties & {\cellcolor{GreenYellow}} 0.36 & {\cellcolor{yellow}} -0.02 & {\cellcolor{yellow}} 0.10 & {\cellcolor{yellow}} 0.27 & {\cellcolor{yellow}} -0.04 & {\cellcolor{yellow}} -0.07 & {\cellcolor{green}} 1.00 \\
\hline
\end{tabular}

\vspace{0.05in}

\begin{tabular}{|l|rrrrrrr|}
\multicolumn{8}{c}{\small{Cardano}} \\
\hline
 & Gini & HHI & CR3 & Entropy & NC & T0.33 & \# Parties \\
 \hline
Gini & {\cellcolor{green}} 1.00 & {\cellcolor{green}} 1.00 & {\cellcolor{green}} 0.99 & {\cellcolor{green}} 1.00 & {\cellcolor{green}} 0.98 & {\cellcolor{green}} 0.97 & {\cellcolor{green}} 0.92 \\
HHI & {\cellcolor{green}} 1.00 & {\cellcolor{green}} 1.00 & {\cellcolor{green}} 1.00 & {\cellcolor{green}} 1.00 & {\cellcolor{green}} 0.98 & {\cellcolor{green}} 0.97 & {\cellcolor{green}} 0.92 \\
CR3 & {\cellcolor{green}} 0.99 & {\cellcolor{green}} 1.00 & {\cellcolor{green}} 1.00 & {\cellcolor{green}} 1.00 & {\cellcolor{green}} 0.98 & {\cellcolor{green}} 0.97 & {\cellcolor{green}} 0.92 \\
Entropy & {\cellcolor{green}} 1.00 & {\cellcolor{green}} 1.00 & {\cellcolor{green}} 1.00 & {\cellcolor{green}} 1.00 & {\cellcolor{green}} 0.98 & {\cellcolor{green}} 0.97 & {\cellcolor{green}} 0.92 \\
NC & {\cellcolor{green}} 0.98 & {\cellcolor{green}} 0.98 & {\cellcolor{green}} 0.98 & {\cellcolor{green}} 0.98 & {\cellcolor{green}} 1.00 & {\cellcolor{green}} 0.99 & {\cellcolor{green}} 0.90 \\
T0.33 & {\cellcolor{green}} 0.97 & {\cellcolor{green}} 0.97 & {\cellcolor{green}} 0.97 & {\cellcolor{green}} 0.97 & {\cellcolor{green}} 0.99 & {\cellcolor{green}} 1.00 & {\cellcolor{LimeGreen}} 0.89 \\
\# Parties & {\cellcolor{green}} 0.92 & {\cellcolor{green}} 0.92 & {\cellcolor{green}} 0.92 & {\cellcolor{green}} 0.92 & {\cellcolor{green}} 0.90 & {\cellcolor{LimeGreen}} 0.89 & {\cellcolor{green}} 1.00 \\
\hline
\end{tabular}

\vspace{0.05in}

\begin{tabular}{|l|rrrrrrr|}
\multicolumn{8}{c}{\small{Ethereum PoW}} \\
\hline
 & Gini & HHI & CR3 & Entropy & NC & T0.33 & \# Parties \\
 \hline
Gini & {\cellcolor{green}} 1.00 & {\cellcolor{LimeGreen}} 0.81 & {\cellcolor{LimeGreen}} 0.71 & {\cellcolor{green}} 0.98 & {\cellcolor{GreenYellow}} 0.43 & {\cellcolor{yellow}} 0.23 & {\cellcolor{pink}} -0.34 \\
HHI & {\cellcolor{LimeGreen}} 0.81 & {\cellcolor{green}} 1.00 & {\cellcolor{green}} 0.96 & {\cellcolor{LimeGreen}} 0.87 & {\cellcolor{LimeGreen}} 0.73 & {\cellcolor{yellow}} 0.23 & {\cellcolor{pink}} -0.38 \\
CR3 & {\cellcolor{LimeGreen}} 0.71 & {\cellcolor{green}} 0.96 & {\cellcolor{green}} 1.00 & {\cellcolor{LimeGreen}} 0.78 & {\cellcolor{LimeGreen}} 0.75 & {\cellcolor{yellow}} 0.23 & {\cellcolor{pink}} -0.46 \\
Entropy & {\cellcolor{green}} 0.98 & {\cellcolor{LimeGreen}} 0.87 & {\cellcolor{LimeGreen}} 0.78 & {\cellcolor{green}} 1.00 & {\cellcolor{GreenYellow}} 0.47 & {\cellcolor{yellow}} 0.23 & {\cellcolor{pink}} -0.40 \\
NC & {\cellcolor{GreenYellow}} 0.43 & {\cellcolor{LimeGreen}} 0.73 & {\cellcolor{LimeGreen}} 0.75 & {\cellcolor{GreenYellow}} 0.47 & {\cellcolor{green}} 1.00 & {\cellcolor{yellow}} 0.28 & {\cellcolor{yellow}} -0.05 \\
T0.33 & {\cellcolor{yellow}} 0.23 & {\cellcolor{yellow}} 0.23 & {\cellcolor{yellow}} 0.23 & {\cellcolor{yellow}} 0.23 & {\cellcolor{yellow}} 0.28 & {\cellcolor{green}} 1.00 & {\cellcolor{yellow}} 0.23 \\
\# Parties & {\cellcolor{pink}} -0.34 & {\cellcolor{pink}} -0.38 & {\cellcolor{pink}} -0.46 & {\cellcolor{pink}} -0.40 & {\cellcolor{yellow}} -0.05 & {\cellcolor{yellow}} 0.23 & {\cellcolor{green}} 1.00 \\
\hline
\end{tabular}

\vspace{0.05in}
\begin{tabular}{|l|rrrrrrr|}
\multicolumn{8}{c}{\small{Ethereum PoS}} \\
\hline
 & Gini & HHI & CR3 & Entropy & NC & T0.33 & \# Parties \\
 \hline
Gini & {\cellcolor{green}} 1.00 & {\cellcolor{lime}} 0.66 & {\cellcolor{LimeGreen}} 0.71 & {\cellcolor{LimeGreen}} 0.79 & {\cellcolor{GreenYellow}} 0.39 & {\cellcolor{yellow}} 0.29 & {\cellcolor{green}} 0.98 \\
HHI & {\cellcolor{lime}} 0.66 & {\cellcolor{green}} 1.00 & {\cellcolor{green}} 0.91 & {\cellcolor{green}} 0.94 & {\cellcolor{lime}} 0.61 & {\cellcolor{GreenYellow}} 0.47 & {\cellcolor{lime}} 0.59 \\
CR3 & {\cellcolor{LimeGreen}} 0.71 & {\cellcolor{green}} 0.91 & {\cellcolor{green}} 1.00 & {\cellcolor{LimeGreen}} 0.89 & {\cellcolor{GreenYellow}} 0.46 & {\cellcolor{yellow}} 0.18 & {\cellcolor{LimeGreen}} 0.71 \\
Entropy & {\cellcolor{LimeGreen}} 0.79 & {\cellcolor{green}} 0.94 & {\cellcolor{LimeGreen}} 0.89 & {\cellcolor{green}} 1.00 & {\cellcolor{lime}} 0.55 & {\cellcolor{GreenYellow}} 0.47 & {\cellcolor{LimeGreen}} 0.70 \\
NC & {\cellcolor{GreenYellow}} 0.39 & {\cellcolor{lime}} 0.61 & {\cellcolor{GreenYellow}} 0.46 & {\cellcolor{lime}} 0.55 & {\cellcolor{green}} 1.00 & {\cellcolor{yellow}} 0.27 & {\cellcolor{GreenYellow}} 0.31 \\
T0.33 & {\cellcolor{yellow}} 0.29 & {\cellcolor{GreenYellow}} 0.47 & {\cellcolor{yellow}} 0.18 & {\cellcolor{GreenYellow}} 0.47 & {\cellcolor{yellow}} 0.27 & {\cellcolor{green}} 1.00 & {\cellcolor{yellow}} 0.14 \\
\# Parties & {\cellcolor{green}} 0.98 & {\cellcolor{lime}} 0.59 & {\cellcolor{LimeGreen}} 0.71 & {\cellcolor{LimeGreen}} 0.70 & {\cellcolor{GreenYellow}} 0.31 & {\cellcolor{yellow}} 0.14 & {\cellcolor{green}} 1.00 \\
\hline
\end{tabular}

\vspace{0.05in}
\begin{tabular}{|l|rrrrrrr|}
\multicolumn{8}{c}{\small{Litecoin}} \\
\hline
 & Gini & HHI & CR3 & Entropy & NC & T0.33 & \# Parties \\
 \hline
Gini & {\cellcolor{green}} 1.00 & {\cellcolor{LimeGreen}} 0.89 & {\cellcolor{LimeGreen}} 0.89 & {\cellcolor{green}} 0.99 & {\cellcolor{lime}} 0.66 & {\cellcolor{GreenYellow}} 0.47 & {\cellcolor{lime}} 0.52 \\
HHI & {\cellcolor{LimeGreen}} 0.89 & {\cellcolor{green}} 1.00 & {\cellcolor{green}} 0.97 & {\cellcolor{LimeGreen}} 0.90 & {\cellcolor{LimeGreen}} 0.82 & {\cellcolor{lime}} 0.69 & {\cellcolor{yellow}} 0.30 \\
CR3 & {\cellcolor{LimeGreen}} 0.89 & {\cellcolor{green}} 0.97 & {\cellcolor{green}} 1.00 & {\cellcolor{LimeGreen}} 0.89 & {\cellcolor{LimeGreen}} 0.81 & {\cellcolor{lime}} 0.63 & {\cellcolor{GreenYellow}} 0.36 \\
Entropy & {\cellcolor{green}} 0.99 & {\cellcolor{LimeGreen}} 0.90 & {\cellcolor{LimeGreen}} 0.89 & {\cellcolor{green}} 1.00 & {\cellcolor{lime}} 0.64 & {\cellcolor{GreenYellow}} 0.45 & {\cellcolor{lime}} 0.50 \\
NC & {\cellcolor{lime}} 0.66 & {\cellcolor{LimeGreen}} 0.82 & {\cellcolor{LimeGreen}} 0.81 & {\cellcolor{lime}} 0.64 & {\cellcolor{green}} 1.00 & {\cellcolor{LimeGreen}} 0.76 & {\cellcolor{yellow}} 0.20 \\
T0.33 & {\cellcolor{GreenYellow}} 0.47 & {\cellcolor{lime}} 0.69 & {\cellcolor{lime}} 0.63 & {\cellcolor{GreenYellow}} 0.45 & {\cellcolor{LimeGreen}} 0.76 & {\cellcolor{green}} 1.00 & {\cellcolor{yellow}} 0.06 \\
\# Parties & {\cellcolor{lime}} 0.52 & {\cellcolor{yellow}} 0.30 & {\cellcolor{GreenYellow}} 0.36 & {\cellcolor{lime}} 0.50 & {\cellcolor{yellow}} 0.20 & {\cellcolor{yellow}} 0.06 & {\cellcolor{green}} 1.00 \\
\hline
\end{tabular}

\vspace{0.05in}
\begin{tabular}{|l|rrrrrrr|}
\multicolumn{8}{c}{\small{Tezos}} \\
\hline
 & Gini & HHI & CR3 & Entropy & NC & T0.33 & \# Parties \\
 \hline
Gini & {\cellcolor{green}} 1.00 & {\cellcolor{yellow}} 0.06 & {\cellcolor{yellow}} 0.19 & {\cellcolor{lime}} 0.53 & {\cellcolor{lime}} 0.56 & {\cellcolor{yellow}} -0.01 & {\cellcolor{LimeGreen}} 0.73 \\
HHI & {\cellcolor{yellow}} 0.06 & {\cellcolor{green}} 1.00 & {\cellcolor{green}} 0.92 & {\cellcolor{LimeGreen}} 0.77 & {\cellcolor{lime}} 0.64 & {\cellcolor{green}} 0.92 & {\cellcolor{yellow}} -0.08 \\
CR3 & {\cellcolor{yellow}} 0.19 & {\cellcolor{green}} 0.92 & {\cellcolor{green}} 1.00 & {\cellcolor{LimeGreen}} 0.88 & {\cellcolor{LimeGreen}} 0.83 & {\cellcolor{green}} 0.92 & {\cellcolor{yellow}} 0.12 \\
Entropy & {\cellcolor{lime}} 0.53 & {\cellcolor{LimeGreen}} 0.77 & {\cellcolor{LimeGreen}} 0.88 & {\cellcolor{green}} 1.00 & {\cellcolor{green}} 0.91 & {\cellcolor{LimeGreen}} 0.72 & {\cellcolor{yellow}} 0.29 \\
NC & {\cellcolor{lime}} 0.56 & {\cellcolor{lime}} 0.64 & {\cellcolor{LimeGreen}} 0.83 & {\cellcolor{green}} 0.91 & {\cellcolor{green}} 1.00 & {\cellcolor{lime}} 0.62 & {\cellcolor{GreenYellow}} 0.43 \\
T0.33 & {\cellcolor{yellow}} -0.01 & {\cellcolor{green}} 0.92 & {\cellcolor{green}} 0.92 & {\cellcolor{LimeGreen}} 0.72 & {\cellcolor{lime}} 0.62 & {\cellcolor{green}} 1.00 & {\cellcolor{yellow}} -0.14 \\
\# Parties & {\cellcolor{LimeGreen}} 0.73 & {\cellcolor{yellow}} -0.08 & {\cellcolor{yellow}} 0.12 & {\cellcolor{yellow}} 0.29 & {\cellcolor{GreenYellow}} 0.43 & {\cellcolor{yellow}} -0.14 & {\cellcolor{green}} 1.00 \\
\hline
\end{tabular}

\caption{Consensus layer correlations between decentralization metrics.}
\label{tab:consensus-correlations-all}
\end{table*}

\begin{table*}[t]
\centering
\begin{tabular}{|l|rrrrrrr|}
\multicolumn{8}{c}{Bitcoin} \\
\hline
 & Gini & HHI & CR3 & Entropy & NC & T0.33 & \# Parties \\
 \hline
Gini & {\cellcolor{green}} 1.00 & {\cellcolor{pink}} -0.43 & {\cellcolor{yellow}} -0.21 & {\cellcolor{orange}} -0.60 & {\cellcolor{orange}} -0.62 & {\cellcolor{orange}} -0.68 & {\cellcolor{orange}} -0.63 \\
HHI & {\cellcolor{pink}} -0.43 & {\cellcolor{green}} 1.00 & {\cellcolor{LimeGreen}} 0.87 & {\cellcolor{yellow}} -0.03 & {\cellcolor{yellow}} 0.15 & {\cellcolor{yellow}} 0.06 & {\cellcolor{yellow}} -0.09 \\
CR3 & {\cellcolor{yellow}} -0.21 & {\cellcolor{LimeGreen}} 0.87 & {\cellcolor{green}} 1.00 & {\cellcolor{pink}} -0.33 & {\cellcolor{yellow}} -0.08 & {\cellcolor{yellow}} -0.22 & {\cellcolor{pink}} -0.36 \\
Entropy & {\cellcolor{orange}} -0.60 & {\cellcolor{yellow}} -0.03 & {\cellcolor{pink}} -0.33 & {\cellcolor{green}} 1.00 & {\cellcolor{LimeGreen}} 0.86 & {\cellcolor{green}} 0.94 & {\cellcolor{green}} 0.98 \\
NC & {\cellcolor{orange}} -0.62 & {\cellcolor{yellow}} 0.15 & {\cellcolor{yellow}} -0.08 & {\cellcolor{LimeGreen}} 0.86 & {\cellcolor{green}} 1.00 & {\cellcolor{LimeGreen}} 0.88 & {\cellcolor{LimeGreen}} 0.82 \\
T0.33 & {\cellcolor{orange}} -0.68 & {\cellcolor{yellow}} 0.06 & {\cellcolor{yellow}} -0.22 & {\cellcolor{green}} 0.94 & {\cellcolor{LimeGreen}} 0.88 & {\cellcolor{green}} 1.00 & {\cellcolor{green}} 0.93 \\
\# Parties & {\cellcolor{orange}} -0.63 & {\cellcolor{yellow}} -0.09 & {\cellcolor{pink}} -0.36 & {\cellcolor{green}} 0.98 & {\cellcolor{LimeGreen}} 0.82 & {\cellcolor{green}} 0.93 & {\cellcolor{green}} 1.00 \\
\hline
\end{tabular}

\begin{tabular}{|l|rrrrrrr|}
\multicolumn{8}{c}{Cardano} \\
\hline
 & Gini & HHI & CR3 & Entropy & NC & T0.33 & \# Parties \\
 \hline
Gini & {\cellcolor{green}} 1.00 & {\cellcolor{yellow}} 0.11 & {\cellcolor{yellow}} 0.07 & {\cellcolor{yellow}} 0.13 & {\cellcolor{yellow}} 0.04 & {\cellcolor{yellow}} -0.01 & {\cellcolor{yellow}} -0.19 \\
HHI & {\cellcolor{yellow}} 0.11 & {\cellcolor{green}} 1.00 & {\cellcolor{green}} 0.99 & {\cellcolor{green}} 0.96 & {\cellcolor{LimeGreen}} 0.85 & {\cellcolor{green}} 0.94 & {\cellcolor{LimeGreen}} 0.84 \\
CR3 & {\cellcolor{yellow}} 0.07 & {\cellcolor{green}} 0.99 & {\cellcolor{green}} 1.00 & {\cellcolor{green}} 0.94 & {\cellcolor{LimeGreen}} 0.86 & {\cellcolor{green}} 0.95 & {\cellcolor{LimeGreen}} 0.86 \\
Entropy & {\cellcolor{yellow}} 0.13 & {\cellcolor{green}} 0.96 & {\cellcolor{green}} 0.94 & {\cellcolor{green}} 1.00 & {\cellcolor{green}} 0.91 & {\cellcolor{green}} 0.91 & {\cellcolor{LimeGreen}} 0.85 \\
NC & {\cellcolor{yellow}} 0.04 & {\cellcolor{LimeGreen}} 0.85 & {\cellcolor{LimeGreen}} 0.86 & {\cellcolor{green}} 0.91 & {\cellcolor{green}} 1.00 & {\cellcolor{green}} 0.90 & {\cellcolor{green}} 0.90 \\
T0.33 & {\cellcolor{yellow}} -0.01 & {\cellcolor{green}} 0.94 & {\cellcolor{green}} 0.95 & {\cellcolor{green}} 0.91 & {\cellcolor{green}} 0.90 & {\cellcolor{green}} 1.00 & {\cellcolor{green}} 0.92 \\
\# Parties & {\cellcolor{yellow}} -0.19 & {\cellcolor{LimeGreen}} 0.84 & {\cellcolor{LimeGreen}} 0.86 & {\cellcolor{LimeGreen}} 0.85 & {\cellcolor{green}} 0.90 & {\cellcolor{green}} 0.92 & {\cellcolor{green}} 1.00 \\

\hline
\end{tabular}

\begin{tabular}{|l|rrrrrrr|}
\multicolumn{8}{c}{Ethereum PoW} \\
\hline
 & Gini & HHI & CR3 & Entropy & NC & T0.33 & \# Parties \\
 \hline
Gini & {\cellcolor{green}} 1.00 & {\cellcolor{pink}} -0.50 & {\cellcolor{pink}} -0.49 & {\cellcolor{orange}} -0.67 & {\cellcolor{orange}} -0.60 & {\cellcolor{orange}} -0.65 & {\cellcolor{orange}} -0.61 \\
HHI & {\cellcolor{pink}} -0.50 & {\cellcolor{green}} 1.00 & {\cellcolor{green}} 0.98 & {\cellcolor{LimeGreen}} 0.86 & {\cellcolor{green}} 0.92 & {\cellcolor{LimeGreen}} 0.87 & {\cellcolor{yellow}} 0.05 \\
CR3 & {\cellcolor{pink}} -0.49 & {\cellcolor{green}} 0.98 & {\cellcolor{green}} 1.00 & {\cellcolor{LimeGreen}} 0.83 & {\cellcolor{LimeGreen}} 0.89 & {\cellcolor{LimeGreen}} 0.83 & {\cellcolor{yellow}} 0.01 \\
Entropy & {\cellcolor{orange}} -0.67 & {\cellcolor{LimeGreen}} 0.86 & {\cellcolor{LimeGreen}} 0.83 & {\cellcolor{green}} 1.00 & {\cellcolor{green}} 0.96 & {\cellcolor{green}} 0.91 & {\cellcolor{GreenYellow}} 0.33 \\
NC & {\cellcolor{orange}} -0.60 & {\cellcolor{green}} 0.92 & {\cellcolor{LimeGreen}} 0.89 & {\cellcolor{green}} 0.96 & {\cellcolor{green}} 1.00 & {\cellcolor{green}} 0.93 & {\cellcolor{yellow}} 0.22 \\
T0.33 & {\cellcolor{orange}} -0.65 & {\cellcolor{LimeGreen}} 0.87 & {\cellcolor{LimeGreen}} 0.83 & {\cellcolor{green}} 0.91 & {\cellcolor{green}} 0.93 & {\cellcolor{green}} 1.00 & {\cellcolor{yellow}} 0.13 \\
\# Parties & {\cellcolor{orange}} -0.61 & {\cellcolor{yellow}} 0.05 & {\cellcolor{yellow}} 0.01 & {\cellcolor{GreenYellow}} 0.33 & {\cellcolor{yellow}} 0.22 & {\cellcolor{yellow}} 0.13 & {\cellcolor{green}} 1.00 \\
\hline
\end{tabular}

\begin{tabular}{|l|rrrrrrr|}
\multicolumn{8}{c}{Ethereum PoS} \\
\hline
 & Gini & HHI & CR3 & Entropy & NC & T0.33 & \# Parties \\
\hline
Gini & {\cellcolor{green}} 1.00 & {\cellcolor{green}} 0.96 & {\cellcolor{green}} 0.96 & {\cellcolor{LimeGreen}} 0.76 & {\cellcolor{lime}} 0.64 & {\cellcolor{LimeGreen}} 0.74 & {\cellcolor{red}} -0.96 \\
HHI & {\cellcolor{green}} 0.96 & {\cellcolor{green}} 1.00 & {\cellcolor{green}} 1.00 & {\cellcolor{lime}} 0.67 & {\cellcolor{lime}} 0.55 & {\cellcolor{lime}} 0.66 & {\cellcolor{red}} -1.00 \\
CR3 & {\cellcolor{green}} 0.96 & {\cellcolor{green}} 1.00 & {\cellcolor{green}} 1.00 & {\cellcolor{lime}} 0.67 & {\cellcolor{lime}} 0.55 & {\cellcolor{lime}} 0.66 & {\cellcolor{red}} -1.00 \\
Entropy & {\cellcolor{LimeGreen}} 0.76 & {\cellcolor{lime}} 0.67 & {\cellcolor{lime}} 0.67 & {\cellcolor{green}} 1.00 & {\cellcolor{green}} 0.96 & {\cellcolor{green}} 0.99 & {\cellcolor{orange}} -0.67 \\
NC & {\cellcolor{lime}} 0.64 & {\cellcolor{lime}} 0.55 & {\cellcolor{lime}} 0.55 & {\cellcolor{green}} 0.96 & {\cellcolor{green}} 1.00 & {\cellcolor{green}} 0.97 & {\cellcolor{orange}} -0.55 \\
T0.33 & {\cellcolor{LimeGreen}} 0.74 & {\cellcolor{lime}} 0.66 & {\cellcolor{lime}} 0.66 & {\cellcolor{green}} 0.99 & {\cellcolor{green}} 0.97 & {\cellcolor{green}} 1.00 & {\cellcolor{orange}} -0.66 \\
\# Parties & {\cellcolor{red}} -0.96 & {\cellcolor{red}} -1.00 & {\cellcolor{red}} -1.00 & {\cellcolor{orange}} -0.67 & {\cellcolor{orange}} -0.55 & {\cellcolor{orange}} -0.66 & {\cellcolor{green}} 1.00 \\
\hline
\end{tabular}

\begin{tabular}{|l|rrrrrrr|}
\multicolumn{8}{c}{Litecoin} \\
\hline
 & Gini & HHI & CR3 & Entropy & NC & T0.33 & \# Parties \\
 \hline
Gini & {\cellcolor{green}} 1.00 & {\cellcolor{yellow}} -0.03 & {\cellcolor{yellow}} -0.01 & {\cellcolor{yellow}} -0.02 & {\cellcolor{yellow}} 0.00 & {\cellcolor{yellow}} -0.23 & {\cellcolor{orange}} -0.69 \\
HHI & {\cellcolor{yellow}} -0.03 & {\cellcolor{green}} 1.00 & {\cellcolor{green}} 0.95 & {\cellcolor{green}} 0.93 & {\cellcolor{LimeGreen}} 0.85 & {\cellcolor{LimeGreen}} 0.88 & {\cellcolor{lime}} 0.61 \\
CR3 & {\cellcolor{yellow}} -0.01 & {\cellcolor{green}} 0.95 & {\cellcolor{green}} 1.00 & {\cellcolor{LimeGreen}} 0.88 & {\cellcolor{LimeGreen}} 0.80 & {\cellcolor{LimeGreen}} 0.73 & {\cellcolor{lime}} 0.58 \\
Entropy & {\cellcolor{yellow}} -0.02 & {\cellcolor{green}} 0.93 & {\cellcolor{LimeGreen}} 0.88 & {\cellcolor{green}} 1.00 & {\cellcolor{green}} 0.95 & {\cellcolor{LimeGreen}} 0.86 & {\cellcolor{lime}} 0.67 \\
NC & {\cellcolor{yellow}} 0.00 & {\cellcolor{LimeGreen}} 0.85 & {\cellcolor{LimeGreen}} 0.80 & {\cellcolor{green}} 0.95 & {\cellcolor{green}} 1.00 & {\cellcolor{LimeGreen}} 0.81 & {\cellcolor{lime}} 0.62 \\
T0.33 & {\cellcolor{yellow}} -0.23 & {\cellcolor{LimeGreen}} 0.88 & {\cellcolor{LimeGreen}} 0.73 & {\cellcolor{LimeGreen}} 0.86 & {\cellcolor{LimeGreen}} 0.81 & {\cellcolor{green}} 1.00 & {\cellcolor{lime}} 0.69 \\
\# Parties & {\cellcolor{orange}} -0.69 & {\cellcolor{lime}} 0.61 & {\cellcolor{lime}} 0.58 & {\cellcolor{lime}} 0.67 & {\cellcolor{lime}} 0.62 & {\cellcolor{lime}} 0.69 & {\cellcolor{green}} 1.00 \\
\hline
\end{tabular}

\begin{tabular}{|l|rrrrrrr|}
\multicolumn{8}{c}{Tezos} \\
\hline
 & Gini & HHI & CR3 & Entropy & NC & T0.33 & \# Parties \\
\hline
Gini & {\cellcolor{green}} 1.00 & {\cellcolor{orange}} -0.54 & {\cellcolor{OrangeRed}} -0.73 & {\cellcolor{pink}} -0.32 & {\cellcolor{pink}} -0.37 & {\cellcolor{pink}} -0.33 & {\cellcolor{orange}} -0.57 \\
HHI & {\cellcolor{orange}} -0.54 & {\cellcolor{green}} 1.00 & {\cellcolor{LimeGreen}} 0.90 & {\cellcolor{green}} 0.92 & {\cellcolor{green}} 0.97 & {\cellcolor{green}} 0.93 & {\cellcolor{LimeGreen}} 0.72 \\
CR3 & {\cellcolor{OrangeRed}} -0.73 & {\cellcolor{LimeGreen}} 0.90 & {\cellcolor{green}} 1.00 & {\cellcolor{LimeGreen}} 0.78 & {\cellcolor{LimeGreen}} 0.82 & {\cellcolor{LimeGreen}} 0.76 & {\cellcolor{LimeGreen}} 0.81 \\
Entropy & {\cellcolor{pink}} -0.32 & {\cellcolor{green}} 0.92 & {\cellcolor{LimeGreen}} 0.78 & {\cellcolor{green}} 1.00 & {\cellcolor{green}} 0.94 & {\cellcolor{green}} 0.95 & {\cellcolor{LimeGreen}} 0.75 \\
NC & {\cellcolor{pink}} -0.37 & {\cellcolor{green}} 0.97 & {\cellcolor{LimeGreen}} 0.82 & {\cellcolor{green}} 0.94 & {\cellcolor{green}} 1.00 & {\cellcolor{green}} 0.95 & {\cellcolor{lime}} 0.65 \\
T0.33 & {\cellcolor{pink}} -0.33 & {\cellcolor{green}} 0.93 & {\cellcolor{LimeGreen}} 0.76 & {\cellcolor{green}} 0.95 & {\cellcolor{green}} 0.95 & {\cellcolor{green}} 1.00 & {\cellcolor{lime}} 0.64 \\
\# Parties & {\cellcolor{orange}} -0.57 & {\cellcolor{LimeGreen}} 0.72 & {\cellcolor{LimeGreen}} 0.81 & {\cellcolor{LimeGreen}} 0.75 & {\cellcolor{lime}} 0.65 & {\cellcolor{lime}} 0.64 & {\cellcolor{green}} 1.00 \\
\hline
\end{tabular}

\caption{Tokenomics layer correlations between decentralization metrics.}
\label{tab:tokenomics-correlations-all}
\end{table*}

\end{document}